# High-field magnetoresistance of microcrystalline and nanocrystalline Ni metal at 3 K and 300 K

I. Bakonyi[1,*], F.D. Czeschka[2,§], L.F. Kiss[1], V.A. Isnaini[1,&], A.T. Krupp[2], K. Palotás[1], S. Zsurzsa[1,#], L. Péter[1]

[1]*Institute for Solid State Physics and Optics, Wigner Research Centre for Physics. H-1121 Budapest, Konkoly-Thege út 29-33, Hungary*

[2]*Walther-Meissner-Institute for Low Temperature Research, Bavarian Academy of Sciences and Humanities. Walther-Meissner-Straße 8, D-85748 Garching, Germany*

---

[*] Corresponding author. E-mail: bakonyi.imre@wigner.hu
[§] Current affiliation: BSH Hausgeräte GmbH, München, Germany
[&] Current address: UIN Sulthan Thaha Saifuddin, Jambi, Indonesia
[#] Current affiliation: Infineon Technologies Cegléd Kft., Cegléd, Hungary

**Abstract** The magnetoresistance (MR) was studied $T$ = 3 K and 300 K for a microcrystalline (μc) Ni foil with grain sizes above 1 micrometer (corresponding to bulk Ni) in magnetic fields up to $H$ = 70 kOe and for a nanocrystalline (nc) Ni foil with an average grain size of about 100 nm up $H$ = 140 kOe in both the longitudinal and transverse configurations. In order to relate the variation of resistivity with magnetic field to the magnetization process, the magnetization isotherms were also measured at both temperatures. At $T$ = 3 K, for the μc-Ni sample with a residual resistivity ratio (RRR) of 331, the field dependence of the resistivity was similar to what was reported previously for high-purity ferromagnets whereas the MR($H$) behavior for the nc-Ni sample with RRR = 9 resembled that what was observed at low temperatures for Ni-based alloys with low impurity concentration. In the magnetically saturated state, the resistivity increased with magnetic field for both samples at $T$ = 3 K and the field dependence was dominated by the ordinary MR due to the Lorentz force acting on the electron trajectories. However, the MR($H$) curves were found to be saturating for μc-Ni and non-saturating for nc-Ni, the difference arising from their very different electron mean free paths. At $T$ = 300 K, the MR($H$) curves of both the μc-Ni and nc-Ni samples were very similar to those known for bulk Ni. After magnetic saturation, the resistivity decreased nearly linearly with magnetic field which behavior is due to the suppression of thermally-induced magnetic disorder with increasing magnetic field and these findings were confronted with previous experimental results and theoretical considerations. The MR($H$) data were analyzed at both temperatures with the help of Kohler plots from which the resistivity anisotropy splitting ($\Delta\rho_{AMR}$) and the anisotropic magnetoresistance (AMR) ratio were derived. It could be demonstrated that at $T$ = 300 K, $\rho(H\rightarrow 0) = \rho(B\rightarrow 0)$ due to the negligible contribution of the ordinary magnetoresistance. In good agreement with our recent room-temperature results on a series of nc-Ni samples with varying grain sizes [V.A. Isnaini et al., *Eur. Phys. J. Plus* **135**, 39 (2020)], the data for the two Ni samples at 3 K and 300 K were found to indicate an approximately linear scaling of $\Delta\rho_{AMR}$ with the zero-field resistivity. This implies that the AMR ratio does not vary significantly with temperature in either microstructural state of Ni metal.

# 1. Introduction

The magnetoresistance (MR) of ferromagnetic metals and alloys has been extensively investigated experimentally in the past [1-6]. Parallel progress in theoretical methods of solid state physics enabled also the calculation of electron transport properties of metallic ferromagnets [7-14]. Although the MR effect in ferromagnets has long been used in various sensor applications [15], after the discovery of the giant magnetoresistance (GMR) effect in magnetic nanostructures [16,17] which opened the way to spintronics, the interest in magnetoresistive sensors further increased. This has boosted a renewed interest in the experimental studies of the MR characteristics of ferromagnetic metals and alloys which constitute an important ingredient in all spintronic devices. In order to get more reliable data on the MR characteristics of bulk ferromagnets, it is of importance, therefore, to carry out new studies on better characterized specimens and eventually under improved and/or extended experimental conditions as done beforehand. These new results may, furthermore, provide a firmer basis for a comparison with the data of theoretical calculations to test the validity of applied models and approximations.

In the spirit of the above considerations, we have turned our attention to studying the electrical transport properties of nanocrystalline (nc) Ni metal. The zero-field electrical resistivity has already been extensively studied for nc-Ni [18-23]. The investigation of the field dependence of the resistivity of nc-Ni has received much less attention [22,24]. In the detailed work by Madduri and Kaul [22], the field dependence was measured in high magnetic fields (up to $H$ = 90 kOe) in a wide temperature range (from $T$ = 1.8 K to 300 K) in a configuration where the measuring current and the applied magnetic field were perpendicular to each other. In a recent paper [24], we have reported and analyzed room-temperature MR data on a series of nc-Ni samples with various grain sizes in a limited magnetic field range (up to $H$ = 9 kOe) with the magnetic field and current either parallel (longitudinal MR = LMR) or perpendicular (transverse MR = TMR) to each other. The zero-field values of the resistivity were determined by extrapolation to $H$ = 0 from the magnetically saturated (monodomain) state for both the LMR and TMR components ($\rho_{Ls}$ and $\rho_{Ts}$, respectively) in a manner as described in Ref. 25. From these data [24], we have then derived parameters characterizing the anisotropic magnetoresistance (AMR) of ferromagnetic metals [4-6,25] by determining the resistivity anisotropy splitting $\Delta\rho_{AMR} = \rho_{Ls} - \rho_{Ts}$ and the anisotropic magnetoresistance ratio defined as AMR = $\Delta\rho_{AMR}/\rho_{is}$. The normalizing factor in the latter expression is the isotropic resistivity $\rho_{is} = (1/3)\rho_{Ls} + (2/3)\rho_{Ts}$, the quantity $\rho_{is}$ being usually very close to the resistivity $\rho_0$ measured in the absence of a magnetic field in the demagnetized state of a ferromagnetic specimen [4].

As an extension of our previous room-temperature study in a limited magnetic field range [24], the major purpose of the present work was to investigate the MR behavior of a nc-Ni sample from Ref. 24 at both low temperature (3 K) and high temperature (300 K) in magnetic fields up to 140 kOe. For purposes of comparison, it was also aimed at measuring for both temperatures the high-field MR characteristics of a coarse-grained, well-annealed and defect-free Ni sample, chosen also from Ref. 24, which actually corresponds to bulk Ni. Since the latter sample is composed of grains with sizes in the micrometer range, we will call it as microcrystalline (μc) Ni. It was expected from this comparison to reveal the significance of the influence of a large amount of grain boundaries on the MR characteristics. In order to relate the variation of the resistivity with magnetic field to the magnetization process, the magnetization isotherms were also measured at both temperatures.

The rationale behind this work is that the magnetoresistance of a nanocrystalline ferromagnetic metal is a very interesting issue. The reason for this is that in the nc state the large density of lattice defects such as grain boundaries represents excess scattering centers for the conduction electrons carrying the current and this leads finally to an increase of the resistivity [26]. Recently, we have summarized [23] all the room-temperature zero-field resistivity data available on nc-Ni and the resistivity of nc-Ni samples was found to be larger than the bulk value [27,28]. Since the resistivity contribution due to the grain boundaries is present at any temperature, the nc state is characterized by a non-negligible residual resistivity as demonstrated, indeed, in several studies on nc-Ni [18-22]. As a consequence, it can be expected that the low-temperature magnetoresistance behavior can also be quite different for the microcrystalline (bulk) and nanocrystalline state of a metal for the following reason.

In pure and defect-free metals, at low temperatures where the phonon term to the resistivity is strongly diminished, the residual resistivity becomes very small and the electron mean free path becomes very long (up to the micrometer range). Such a situation occurs when the residual resistivity ratio (RRR) which is the ratio of the room-temperature resistivity and the residual resistivity is sufficiently high. Under these conditions, the electrons can travel long distances in the large (at least micrometer-sized) grains without being scattered by the residual impurities and lattice defects. However, in a magnetic field $H$ applied for measuring the longitudinal magnetoresistance (magnetic field is oriented along the current flow direction), the electrons move on a path turning around the magnetic field lines due to the Lorentz force while travelling towards lower electric potentials. As a consequence, due to the helical electron path, the distance travelled along the electric field lines will be shorter than the electron mean free path between two consecutive scattering events which then manifests itself in an increased resistance, the resistivity increase being the larger, the larger the applied magnetic field. For the TMR configuration (magnetic field perpendicular to the current flow direction), the resistivity increase in a given magnetic field is even stronger than for the LMR configuration. This is because in the TMR configuration the rotation of the electrons due to the Lorentz force happens around a direction perpendicular to the current flow and, therefore, the electrons move backward and forward with respect to the overall drift direction.

The resistivity contribution due to the presence of a magnetic field via the Lorentz force is termed as ordinary magnetoresistance (OMR) which is the only source of resistivity change due to the application of an external magnetic field in normal (non-magnetic) metals [5,6,29]. Since for ferromagnetic metals with sufficiently high RRR value, the OMR term dominates the observed variation of the resistivity with magnetic field at low temperatures, the determination of the ferromagnetic resistivity anisotropy [3-6], i.e., the difference between the longitudinal and transverse magnetoresistance in pure ferromagnetic metals is a difficult task here and usually cannot be accomplished in a straightforward manner [30,31].

However, as we have seen above, if we introduce a large amount of lattice defects, e.g., in the form of grain boundaries, into an otherwise chemically pure metal, this diminishes the electron mean free path even at low temperatures, resulting in a non-negligible residual resistivity (and in a concomitant reduction of RRR). In such cases, the ferromagnetic resistivity anisotropy can be expected to be determined at low temperatures even for a pure ferromagnetic metal in the nc state.

The magnetoresistance of non-magnetic metals has been traditionally analyzed on the basis of Kohler's rule [29] according to which the magnetic-field-induced relative resistivity change $\Delta\rho(H)/\rho_o = [\rho(H) - \rho_o]/\rho_o$ is a function of the ratio $H/\rho_o$ only where $\rho(H)$ is the resistivity $\rho$ in a magnetic field $H$ and $\rho_o$ is the resistivity measured in the absence of a magnetic field. This can be formulated as $\Delta\rho(H)/\rho_o = F(H/\rho_o)$ where the Kohler function $F$ is to be determined empirically from the measured MR($H$) data. The Kohler function $F$ cannot be derived theoretically; nevertheless, it was established to have some general features, e.g., it was found to be dependent on the relative orientation of the magnetic field $H$ and the measuring current and, if the specimen is a single crystal, also on the crystal orientation. The function $F$ is always of saturating form if the magnetic field and the current are parallel (LMR) and, for a single crystal, depending on the topology of the Fermi surface, $F$ is either increasing or saturating if the magnetic field and the current are perpendicular to each other (TMR) [30]. The validity of Kohler's rule has been demonstrated experimentally for most normal (non-magnetic) metals [29,32]: the MR($H$) data from samples with different residual resistivities (due to different concentration of chemical impurities or lattice defects) of a given metal fell on a single curve when displayed according to Kohler's rule in the form of a so-called Kohler plot: $\Delta\rho(H)/\rho_o$ vs. $H/\rho_o$. In the initial studies of the low-temperature magnetoresistance of ferromagnetic metals, the validity of Kohler's rule was questioned until Schwerer and Silcox [30] recognized that in ferromagnets, instead of the magnetic field $H$, the magnetic induction $B = H + 4\pi M_s$ should be used (here $M_s$ is the saturation magnetization of the ferromagnet) in Kohler's rule. The rationale behind this is the fact that, in a ferromagnet, not the externally applied magnetic field $H$, but the magnetic induction $B$ is the effective field acting on the electron trajectories [5,6]. (It is noted that for a correct treatment in the expression of $B$, the demagnetizing field should also be taken into account [5]. However, in our special strip-shaped specimen geometry, the demagnetizing field can be usually neglected besides the other terms for magnetic fields applied in the plane of the thin strip sample [25].)

In the present paper, the MR($H$) data measured at 3 K could be analyzed on the basis of Kohler's rule for both the μc-Ni and nc-Ni samples in spite of the quite different behaviors of their MR($H$) data. Since the μc-Ni sample had very small residual resistivity, the ferromagnetic resistivity anisotropy requiring an extrapolation of the LMR($H$) and TMR($H$) data to zero induction $B$ [5,6,25,30,31] could be determined with some uncertainty only. On the other hand, the residual resistivity was sufficiently large for the nc-Ni sample which enabled a reliable extrapolation of the LMR($H$) and TMR($H$) data to $B = 0$ by using the Kohler plot and, thus, it was possible to derive the low-temperature ferromagnetic resistivity anisotropy fairly accurately.

In contrast to the case of $T = 3$ K, in the high-temperature regime such as at 300 K, the phonon contribution to the resistivity is the dominant term and, thus, the mean free path is strongly reduced for any metal even in a pure, coarse-grained and defect-free sample. Therefore, at room temperature, the OMR term becomes negligible even in pure metals [33,34]. In this case, the partial suppression of the thermally-induced spin disorder by the increasing magnetic field is the only source of a field-induced resistivity change in the magnetically saturated state; this latter term usually exhibits a nearly linear decrease with increasing magnetic field at high temperatures [22,25,33,34]. Therefore, the MR($H$) behaviors of the μc-Ni and nc-Ni samples were found to be very similar at 300 K both qualitatively and quantitatively. At this high temperature, a quantitative analysis with (or even without) Kohler's rule easily enabled the determination of the ferromagnetic resistivity anisotropy which was quite similar for both microstructural states of Ni. The other interest at $T$ = 300 K was in the quantitative characterization of the magnetic-disorder-suppression process with increasing magnetic field which could be performed with high precision for the present samples and the analysis of our experimental data will be confronted with previously suggested theoretical models and experimental results [22,33,34].

In conformity with the above discussion, the paper is organized as follows. In Section 2, the investigated Ni samples, the electronic transport measurement configurations in a magnetic field and the measurement techniques for the magnetoresistance and the magnetization will be presented. The experimental results for the MR behavior and magnetic properties of the μc-Ni and nc-Ni samples studied are presented separately for $T$ = 3 K in Section 3 and for $T$ = 300 K in Section 4. These sections also include a comparison of the results for μc-Ni and nc-Ni at each temperature as well as with previously reported experimental results. Section 5 will be devoted to a discussion of the field dependence of the resistivity in the magnetically saturated (monodomain) state at $T$ = 300 K and a confrontation with previous experimental results and theoretical considerations [22,34]. An analysis of the correlation between $\Delta\rho_{AMR}$ and $\rho_o$ will then be performed in Section 6 similarly as in our recent work on a series of nc-Ni samples with various grain sizes [24]. A summary of the present results will be given in Section 7.

## 2. Experimental
### 2.1 Samples investigated

The results of magnetotransport measurements to be presented here were obtained on a μc-Ni and a nc-Ni thin foil sample. Their room-temperature zero-field resistivity data have been reported in Ref. 23 and the room-temperature magnetoresistance results in magnetic fields up to $H = 9$ kOe in Ref. 24.

The nc-Ni foil (identical with sample #B2 in Refs. 23 and 24) with 9 μm thickness was produced by electrodeposition. Deposition was performed on a polished Ti sheet from which the Ni foil could be mechanically peeled off.

The μc-Ni foil (identical with sample #B5 in Refs. 23 and 24) was obtained by cold rolling a bulk sample of commercial Ni (electrolytic nickel grade) down to a thickness of about 45 μm. After cold rolling, a heat treatment at about 700 °C for 1 h in a protecting hydrogen atmosphere was carried out to release the stresses introduced by the cold-rolling procedure.

According to an energy-dispersive X-ray spectroscopy analysis of the two samples [23], besides a small amount of C and O probably in the form of surface contamination, neither metallic, nor non-metallic impurities could be detected up to a level of about 0.1 at.%.

A detailed structural characterization of these two samples by X-ray diffraction has been reported previously [23]. It was found that the size of the coherently scattering domains (termed as crystallite size) of sample #B5 is in the micrometer range and this justifies to call it as microcrystalline (μc) Ni. As a result of annealing, this is a coarse-grained sample corresponding actually to bulk Ni. The average crystallite size of sample #B2 was about 75 nm as deduced from X-ray diffraction, so it can be considered as a nanocrystalline (nc) Ni sample. It is noted that based on the room-temperature resistivity value of sample #B2 and the reported dependence of resistivity of nanocrystalline Ni on grain size [23], the grain size what could be directly observed by TEM imaging can be estimated to be about 100 nm for sample #B2. In the following, we will use mostly the μc-Ni and nc-Ni notations only for samples #B5 and #B2, respectively.

*2.2 Magnetoresistance and magnetic measurements*

The measurement of the resistivity of thin foils as a function of the externally applied magnetic field $H$ can be conveniently carried out on strip-shaped samples [25]. Therefore, a rectangular strip of about 1 mm wide and about 5 mm long was cut from the Ni foils.

For the magnetoresistance measurements with a four-point probe, spot-welded contacts were attached to the strip-shaped sample. About a dozen contact wires at both ends of the strip served for providing a homogeneous current flow along the strip length. Contact wires were attached at both long edges of the strip to measure the voltage drop along the strip length. The magnetoresistance measurements were performed with the current flowing along the long axis of the strip; the magnetic field was in the plane of the strip and was oriented either parallel (LMR) or perpendicular (TMR) to the current.

By considering the above specified sample shapes and dimensions including the thickness, these samples can be approximated with a general ellipsoid [25] which facilitates the extrapolation of the resistivity to $B = 0$, a procedure necessary under certain circumstances to eliminate the OMR contribution.

For measuring the resistance, a d.c. current with alternating sign and with an amplitude of $I = \pm 100$ µA was applied and the resulting d.c. voltage drop along the strip length was recorded with a nanovoltmeter. The resistance probe could be inserted into the cryostat of a superconducting magnet with magnetic fields up to $H = 140$ kOe. In the cryostat, the sample temperature could be varied from 3 K to 300 K.

The in-plane magnetization isotherms *M(H)* were measured in a SQUID magnetometer at 3 K and 300 K up to $H = 50$ kOe and in a vibrating sample magnetometer (VSM) at 300 K up to $H = 10$ kOe.

## 3. Electrical transport and magnetic properties at $T = 3$ K

*3.1 Zero-field resistivity data*

3.1.1 Zero-field resistivity of μc-Ni

In a previous study [23], we have made careful measurements of the room-temperature zero-field resistivity on thin strip-shaped foil samples of high-purity Ni metals from various sources. For all these samples, with the applied resistivity measurement probe, we could reproduce fairly well (within about ±3 %) the standard reference value for the room-temperature zero-field resistivity $\rho_o$ of bulk Ni at $T = 300$ K ($\rho_o = 7.24$ μΩcm [27,28]). Specifically, the zero-field resistivity of the μc-Ni foil (sample #B5) investigated in the present work was found to be $\rho_o = 7.36 \pm 0.21$ μΩcm at 300 K [23].

In the present work, it was obtained for the zero-field the residual resistivity ratio that RRR = $\rho(300K)/\rho(3K) = 331$ for the same μc-Ni foil, which indicates a fairly high purity and a defect-free state of the sample #B5. For comparison, we note that for the Ni sample which was used to measure the standard resistivity data for bulk Ni [27,28], the residual resistivity ratio was specified as RRR = $\rho(273K)/\rho(4K) = 270$ (in another often referenced bulk Ni resistivity study, the Ni sample had RRR = 310 [35]).

It follows from our experimental RRR value that the zero-field resistivity of the μc-Ni foil at 3 K is $\rho_o = \rho(H=0) = 0.02222$ μΩcm. According to the chemical analysis results mentioned in Section 2.1, metallic impurities if present at all are below the 0.1 at.% level. The most common metallic impurity in Ni is Co and Fe. If we assume that about 0.05 at.% Fe and 0.05 at.% Co impurity is present in our μc-Ni sample, then based on the reported residual resistivity contribution of Co and Fe in Ni [5], we get $\rho_{res} = 0.026$ μΩcm which is very close to our measured $\rho_o$ value.

3.1.2 Zero-field resistivity of nc-Ni

The temperature dependence of the resistivity of nc-Ni measured from 3 K to 300 K in zero magnetic field and normalized to the room-temperature value is displayed in Fig. 1. Very similar behavior was reported on nc-Ni in all previous studies [18-22].

According to the data in Fig. 1, the residual resistivity of the nc-Ni foil is 11 % of the room temperature value which corresponds to a residual resistivity ratio of RRR = 9. The zero-field resistivity of the same nc-Ni foil (sample #B2) at 300 K was determined as $\rho_o =$ 8.78 μΩcm in a previous study [23]. It follows from these data that the zero-field resistivity at $T = 3$ K is $\rho_0 = 0.9739$ μΩcm for the presently investigated nc-Ni sample.

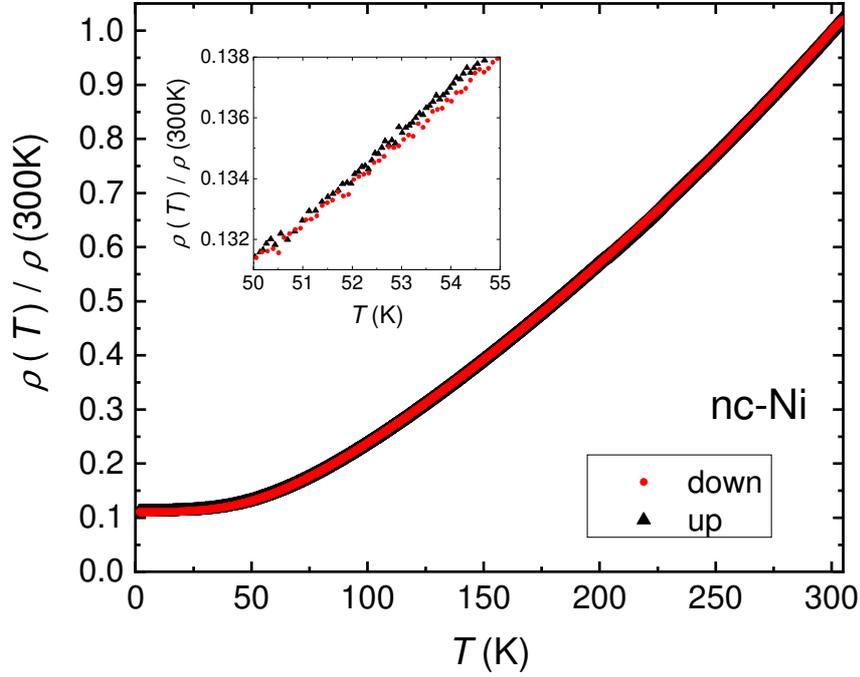

*Figure 1 Temperature dependence of the normalized resistivity measured in zero magnetic field on the nc-Ni foil. The results of temperature scans downwards (•) and upwards (▲) agree with high accuracy as can be seen in the enlarged view of a selected narrow temperature interval in the inset.*

The larger room-temperature resistivity of the nc-Ni sample ($\rho_0 = 8.78$ μΩcm [23]) in comparison with the corresponding value of bulk Ni ($\rho_0 = 7.24$ μΩcm [27,28]) and of the μc-Ni foil ($\rho_0 = 7.36$ μΩcm) is due to the presence of a large density of lattice defects, especially grain boundaries in the nanocrystalline state which represent extra scattering centers for the conduction electrons [26]. As summarized recently [23], a common observation of previous studies on the grain size dependence of the zero-field electrical resistivity of nc-Ni was that the smaller the grain size, the larger the resistivity. This corresponds to expectation on the basis of the increasing contribution to the resistivity by the grain boundaries with decreasing grain size, i.e., with increasing grain boundary volume fraction [23].

Since the grain boundary scattering contribution is present also at low temperatures, the nanocrystalline state exhibits a non-negligible residual resistivity even for very pure metals as observed, indeed, in Fig. 1 for our nc-Ni sample.

*3.2 Magnetoresistance and magnetic properties of µc-Ni at T =3 K*

Figures 2a and 2b show the field dependence of the resistivity for µc-Ni at $T = 3$ K for the LMR and TMR configurations. It can be seen that at this temperature the resistivity minimum at $H = 0$ coincides for the longitudinal and transverse components with a zero-field resistivity $\rho_o = 0.02222$ µΩcm. The data displayed in Fig. 2b demonstrate that, in spite of the relatively large random noise (about 2 % of the measured values), the average values are well reproducible during the various field cycling directions for each configuration and also the $\rho_o$ values for the two components agree very well. This is due to the extremely high stability and accuracy of our measurement setup.

The random noise of the resistivity measurement does not enable to directly see a hysteresis of the MR($H$) data for the µc-Ni sample in Fig. 2b. However, a careful analysis of the TMR results at higher resolution revealed a hysteresis and yielded the peak positions to be at a magnetic field of about $H_p = \pm 85(10)$ Oe.

In order to assess the magnetization process, especially the onset of technical magnetic saturation (i.e., the achievement of the monodomain state), the magnetization curve $M(H)$ of the µc-Ni sample was measured at 3 K by SQUID up to 50 kOe which is shown in Figs. 2c and 2d. Due to the low coercive field of µc-Ni, the SQUID magnetometer could not reveal a magnetic hysteresis (the reason for this is that the power supply of the superconducting magnet of the SQUID is not bipolar), we can only estimate that the coercive field ($H_c$) of the µc-Ni sample can be around 10 Oe.

It can be established from the measured $M(H)$ data that saturation of the magnetization is achieved in a magnetic field of the order of a few kilooersted. At the same time, in the magnetic field range where saturation is reached, there are no specific features in the MR($H$) curves, we may perhaps just notice a change in the slope of the TMR($H$) curve around 0.5 kOe (see Fig. 2b).

It is easy to reveal from Fig. 2a that the µc-Ni sample exhibits a strongly non-linear field dependence of the resistivity at such a low temperature. The MR($H$) data indicate that TMR > LMR in the whole range of magnetic fields investigated and both components tend to show a saturation character for high magnetic fields. These features correspond to the discussion in the Introduction about the magnetoresistance behavior of pure ferromagnetic metals: the low residual resistivity (high RRR ratio) results in an extremely large electron mean free path owing to which the electron trajectories are strongly deflected by the magnetic field, i.e., a large OMR effect occurs and a transverse magnetic field causes a larger effect than a longitudinal field. Very similar low-temperature MR($H$) curves were reported for high-purity Ni both for single-crystal [36,37] and polycrystalline [38] samples although no attempt was made in those studies to determine the magnitude of the anisotropic magnetoresistance effect.

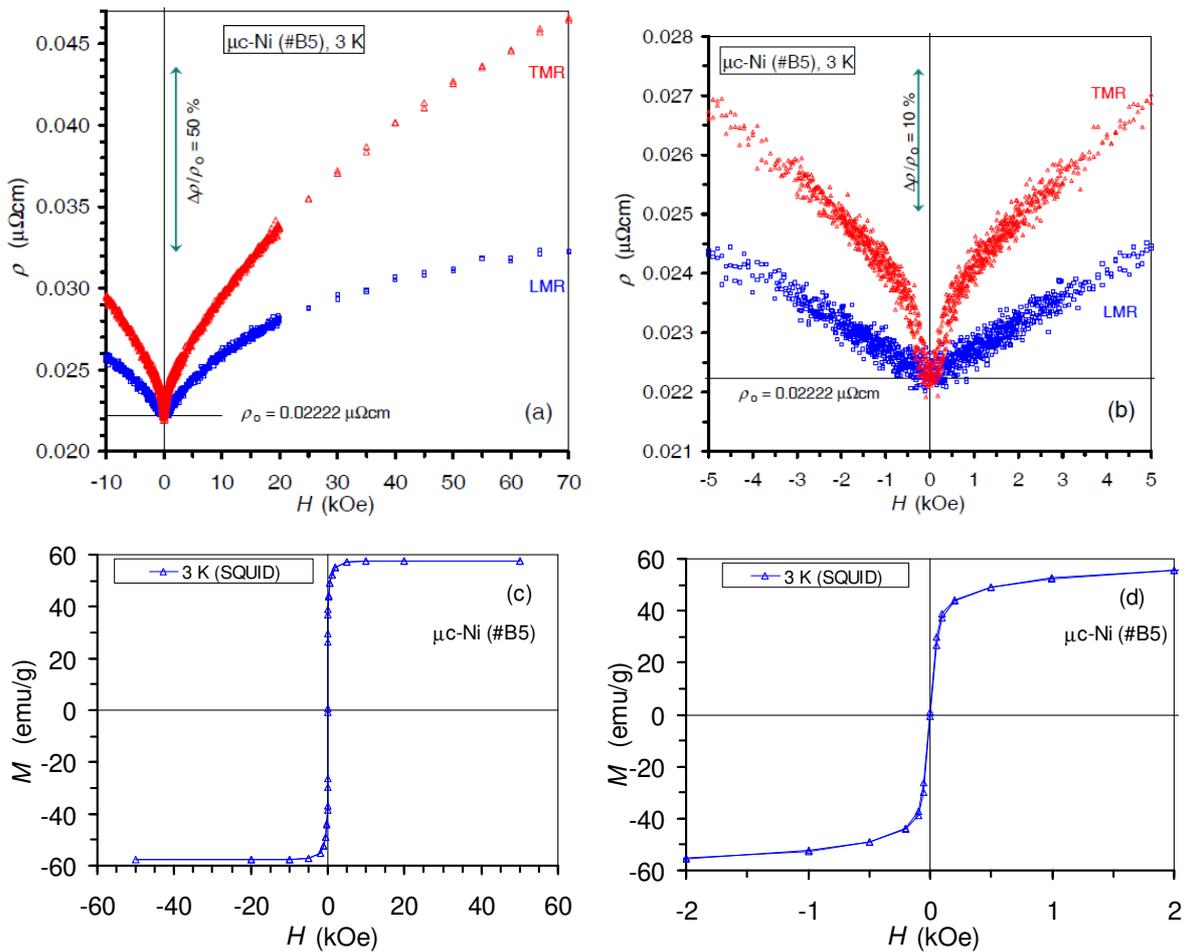

***Figure 2***  *(a) Field-dependence of the resistivity $\rho$ at T = 3 K for $\mu$c-Ni with magnetic field orientations as indicated (LMR, TMR) in the magnetic field range from -10 kOe to +70 kOe. (b) Enlarged version of the resistivity data at low magnetic fields (-5 kOe $\leq H \leq$ + 5 kOe). (c) High-field magnetization curve M(H) at 3 K for the $\mu$c-Ni sample. (d) Low-field magnetization curve M(H) at 3 K for the $\mu$c-Ni sample. Note: the vertical double-headed arrows in (a) and (b) indicate the specified relative resistivity change $\delta\rho/\rho_o$.*

We will analyze the data on the basis of the Kohler plots. As suggested by Schwerer and Silcox [30], in applying Kohler's rule for ferromagnets, the magnetic field $H$ should be replaced by the magnetic induction $B = H + 4\pi M_s$. Accordingly, Kohler's rule for a ferromagnet will take the form $\Delta\rho(B)/\rho(B=0) = [\rho(B) - \rho(B=0)]/\rho(B=0) = F[B/\rho(B=0)]$ where we have now replaced the zero-field resistivity $\rho_o = \rho(H=0)$ with $\rho(B=0)$ to emphasize that the resistivity at zero induction is the normalizing factor for ferromagnets. Kohler's rule can be transformed also into the following form: $\rho(B)/\rho(B=0) = 1 + F[B/\rho(B=0)]$. This implies that if we display $\rho(B)/\rho(B=0)$ as a function of $B/\rho(B=0)$ which is the Kohler plot, then the Kohler function $F[B/\rho(B=0)]$ should extrapolate to 1 when $B \to 0$.

In order to carry out the data analysis on the basis of the Kohler plot, we should know the value of $\rho(B=0)$. To find the experimentally unattainable zero-induction resistivity $\rho(B=0)$, the procedure is the following. We display the $\rho(B)/\rho(B=0)$ data as a function of $B/\rho(B=0)$ by taking first $\rho(B=0)$ equal to the zero-field resistivity $\rho(H=0)$. Since the form of the Kohler function $F$ is not known, we fit the data with a polynomial function by using the constraint that the Kohler function $F[B/\rho(B=0)]$ should extrapolate to 1 when $B \to 0$. It was found that a fourth-order polynomial is sufficient since going to higher orders, the normalized fit quality parameter ($R^2$) provided by the Excel fitting program as the square of the Pearson product moment correlation coefficient improved only insignificantly. By finely tuning the value of $\rho(B=0)$, after a few trials we can find the appropriate value which yields a fit with good quality ($R^2 > 0.99$). This is done separately for both the LMR and TMR components.

Evidently, data in the magnetically saturated state should only be taken into account in the Kohler plot analysis. We have found that for the µc-Ni sample the resistivity data for $H \geq$ 3 kOe correspond to this criterion by considering the measured $M(H)$ isotherm (see Figs. 2c and 2d); by restricting data to higher fields did not improve the fit quality in the data analysis.

The Kohler plot with the fitting functions (solid lines) obtained at maximized fit quality parameter $R^2$ is shown in Fig. 3 for the LMR($H$) and TMR($H$) data measured at $T = 3$ K for the µc-Ni sample.

As a result of the fitting procedure, we obtained the zero-induction resistivities $\rho(B=0)$ for both the LMR and TMR components. These values are specified in Fig. 3 in the left upper corner text box and they correspond to the previously mentioned saturation resistivities [25]: $\rho_L(B=0) = \rho_{Ls}$ and $\rho_T(B=0) = \rho_{Ts}$. From these fitted zero-induction resistivities, we can then derive the isotropic resistivity $\rho_{is} = (1/3)\,\rho_{Ls} + (2/3)\,\rho_{Ts}$, the resistivity anisotropy splitting $\Delta\rho_{AMR} = \rho_{Ls} - \rho_{Ts}$ and the AMR ratio = $\Delta\rho_{AMR}/\rho_{is}$ [25] which values are also specified in the same textbox.

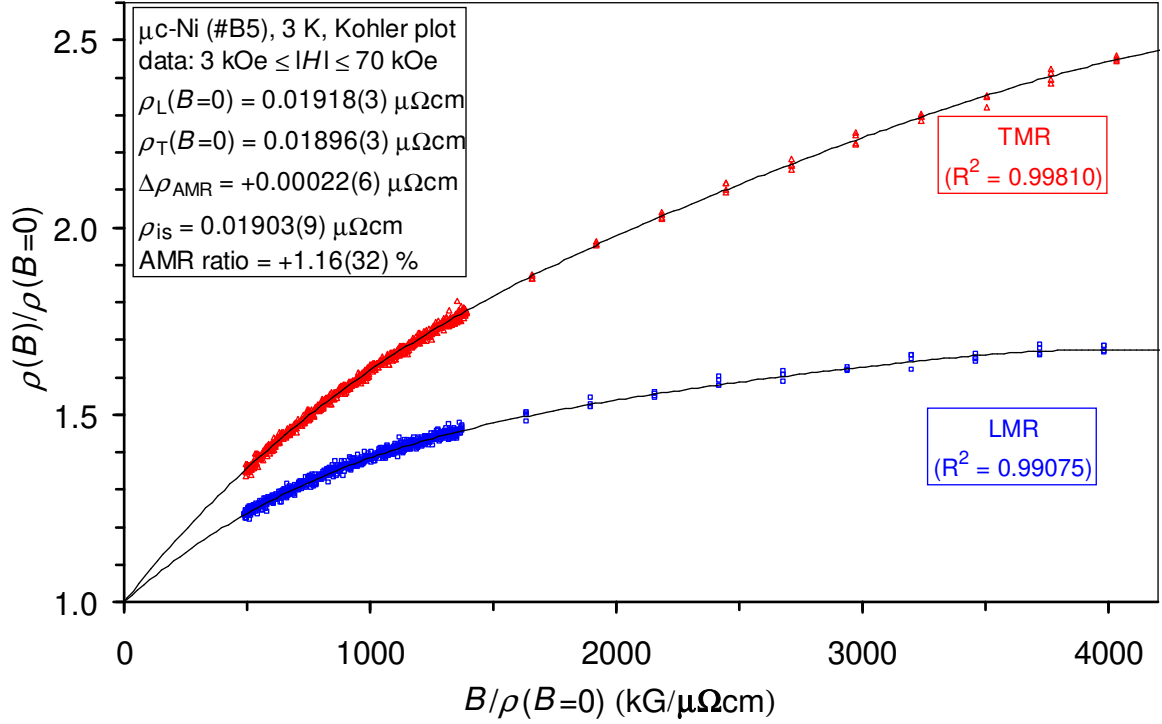

***Figure 3*** *Kohler plot $\rho(B)/\rho(B=0)$ vs. $B/\rho(B=0)$ at $T = 3$ K for µc-Ni with magnetic field orientations as indicated (LMR, TMR) for the MR(H) data in the magnetic field range 3 kOe ≤ |H| ≤ 70 kOe. The experimental data are the symbols (squares: LMR; triangles: TMR), the solid lines are the fourth-order polynomial fitting functions providing an empirical analytical form of the Kohler function F. The fitting parameter values were obtained at the maximum values of the normalized fit quality parameter $R^2$ specified in the text boxes attached to the LMR and TMR curves. The extrapolated zero-induction resistivities $\rho(B=0)$ for the LMR and TMR components, the isotropic resistivity $\rho_{is}$ as well as the derived resistivity anisotropy parameters ($\Delta\rho_{AMR}$ and AMR ratio) are given in the text box in the upper left corner (see text for more details).*

The errors of the fit parameters $\rho_L(B=0)$ and $\rho_T(B=0)$ specified in the brackets after the parameter values correspond to the range within which the $R^2$ value had a maximum and the errors given for the derived parameters were deduced from these uncertainties of the two fit parameters. The resulting final absolute error for the magnitude of the AMR ratio was then ±0.32 %. However, the actual error is certainly larger due to the uncertainty of the extrapolation to $B = 0$ in lack of an exact knowledge of the Kohler function $F$. Since the resistivity $\rho(B=0)$ is fairly small and the minimum value of the induction is $B = H_s + 4\pi M_s =$ 9.4 kG (where $H_s = 3$ kOe, i.e., the magnetic field considered as necessary for achieving the monodomain state), the extrapolation to $B = 0$ should be made from a fairly high value on the $B/\rho(B=0)$ axis.

This difficulty was already noticed also by Schwerer and Silcox [31] so they used Ni samples containing controlled small amounts of impurities resulting in higher resistivities to get data also close to $B/\rho(B=0) = 0$, but here we have a single sample only. We will see in the next section that the difficulty of the extrapolation to $B = 0$ at $T = 3$ K is lifted for the nc-Ni sample with much higher residual resistivity.

Another source of uncertainty is the relatively large scatter of the LMR($H$) data for which the fit quality parameter $R^2$ was definitely smaller than for the TMR($H$) data as can be seen in Fig. 3. As a consequence, the fitting function with the maximum $R^2 = 0.99075$ value yielding the specified $\rho_L(B=0) = 0.01918$ μΩcm value (from which 1.16 % was deduced for the AMR ratio) did show some systematic deviation from the data when viewed at larger magnification. Therefore, we have analyzed the LMR($H$) data also in a manner to get an optimum fit to the data by visual inspection (i.e., without apparent systematic deviation). This was achieved at the fit parameter value of about $\rho_L(B=0) = 0.01960$ μΩcm with $R^2 = 0.99057$ and this resulted in an AMR ratio of 3.34 %. Evidently, the values of the fit quality parameter $R^2$ of the two approaches are so close to each other that both fitted parameter values of $\rho_L(B=0)$ can be considered as of equal reliability. By taking formally the average of the above two AMR ratio values, we end up with an AMR ratio of (2.25 ± 1.10) % for the μc-Ni sample at $T = 3$ K.

In order to check the reliability of the above determined resistivity anisotropy parameter values, we have also made a further test. The measured resistivity data were displayed against the magnetic induction $B$ as shown in Fig. 4. A similar fourth-order polynomial was fitted to the experimental data as in the Kohler plot (Fig. 3) with the same fixed $\rho(B=0)$ values as derived from the Kohler plot analysis. It can be seen that the fit quality, both in terms of $R^2$ as well as a proper data fitting, is the same as in the Kohler plot of Fig. 3 (for the LMR($H$) data, this was found to be valid for both above obtained $\rho_L(B=0)$ values). Furthermore, when fitting the $\rho$ vs. $B$ data without fixing the value of $\rho(B=0)$ to that of the Kohler plot analysis, but rather allowing it as a free parameter, the fit quality parameter $R^2$ remained the same up to the fifth digit. At the same time, the fitted $\rho_T(B=0)$ value also remained the same as obtained from the Kohler plot and $\rho_L(B=0)$ changed by 1 in the last (fifth) digit due to the more noisy LMR($H$) data. This justifies that the forced fit used in evaluating the Kohler diagram is also appropriate for describing the experimental $\rho$ vs. $B$ data.

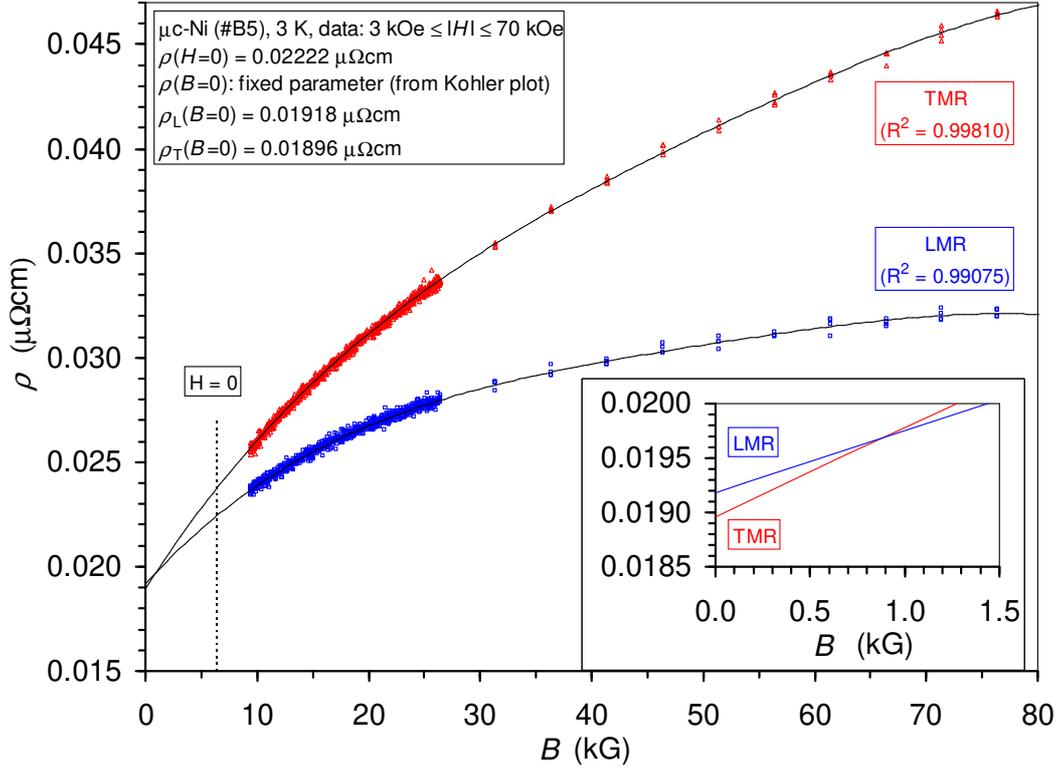

***Figure 4*** *Resistivity $\rho$ vs. B at T = 3 K for μc-Ni with magnetic field orientations as indicated (LMR, TMR) for the MR(H) data in the magnetic field range 3 kOe $\leq |H| \leq$ 70 kOe. The experimental data are the symbols (squares: LMR; triangles: TMR), the solid lines are the fourth-order polynomial fitting functions providing a proper description of the evolution of the measured resistivity data with magnetic induction. During fitting, the zero-induction resistivities $\rho(B=0)$ for the LMR and TMR components were fixed at the values taken from the Kohler plot (Fig. 3). The vertical dashed line indicates the value of the magnetic induction B at H = 0. The inset shows the data close to B = 0 in order to better resolve the relation $\rho_{LMR}(B=0) > \rho_{TMR}(B=0)$ here.*

We can also see in Fig. 4 that if we had extrapolated the resistivity to $H = 0$ ($B = 6.4$ kG), the position of which is indicated by the vertical dashed line, we would have obtained an opposite relation for the extrapolated LMR and TMR components [$\rho_{LMR}(H=0)$ <

$\rho_{TMR}(H=0)$] which would have resulted in a negative AMR ratio (AMR < 0). This is a clear indication that for these data, indeed, the Kohler plot analysis [$\rho(B)/\rho(B=0)$ vs. $B/\rho(B=0)$] is required to determine the true magnetotransport parameters. This also implies, at the same time, that the observed field-induced resistivity change is dominated by the OMR effect.

The coefficients of the fitted fourth-order polynomial functions of the LMR and TMR components of the resistivity for the µc-Ni sample as a function of the magnetic induction will be summarized in Table I in subsection 3.4.

### 3.3 Magnetoresistance and magnetic properties of nc-Ni at T =3 K

Figures 5a and 5b show the field dependence of the resistivity of nc-Ni for both measurement configurations (LMR, TMR) at 3 K. Similarly to the data on the µc-Ni sample (Fig. 2b), the resistivity values for the nc-Ni sample are also well reproducible during the various field cycling directions for each configuration. The $\rho_o$ values for the two components also agree very well by noting that for the less noisy data of the TMR component (Fig. 5b) which reveal a clear hysteresis, $\rho_o$ is identified as the peak value of the resistivity $H_p$, i.e., $\rho_o = \rho(H_p)$. The peak positions of the TMR(H) curves are at $H_p = \pm 140$ Oe.

The $M(H)$ curves of the nc-Ni sample as measured by the SQUID are displayed in Figs. 5c and 5d. The low-field $M(H)$ curves (Fig. 5d) also reveal hysteresis with a coercive field of about 45 Oe.

As to the field dependence of the magnetization, the $M(H)$ data in Fig. 5d show that technical magnetic saturation is certainly achieved in a few kilooersted of magnetic field. In the same magnetic field range, the MR(H) curves show a change in the field-evolution trend.

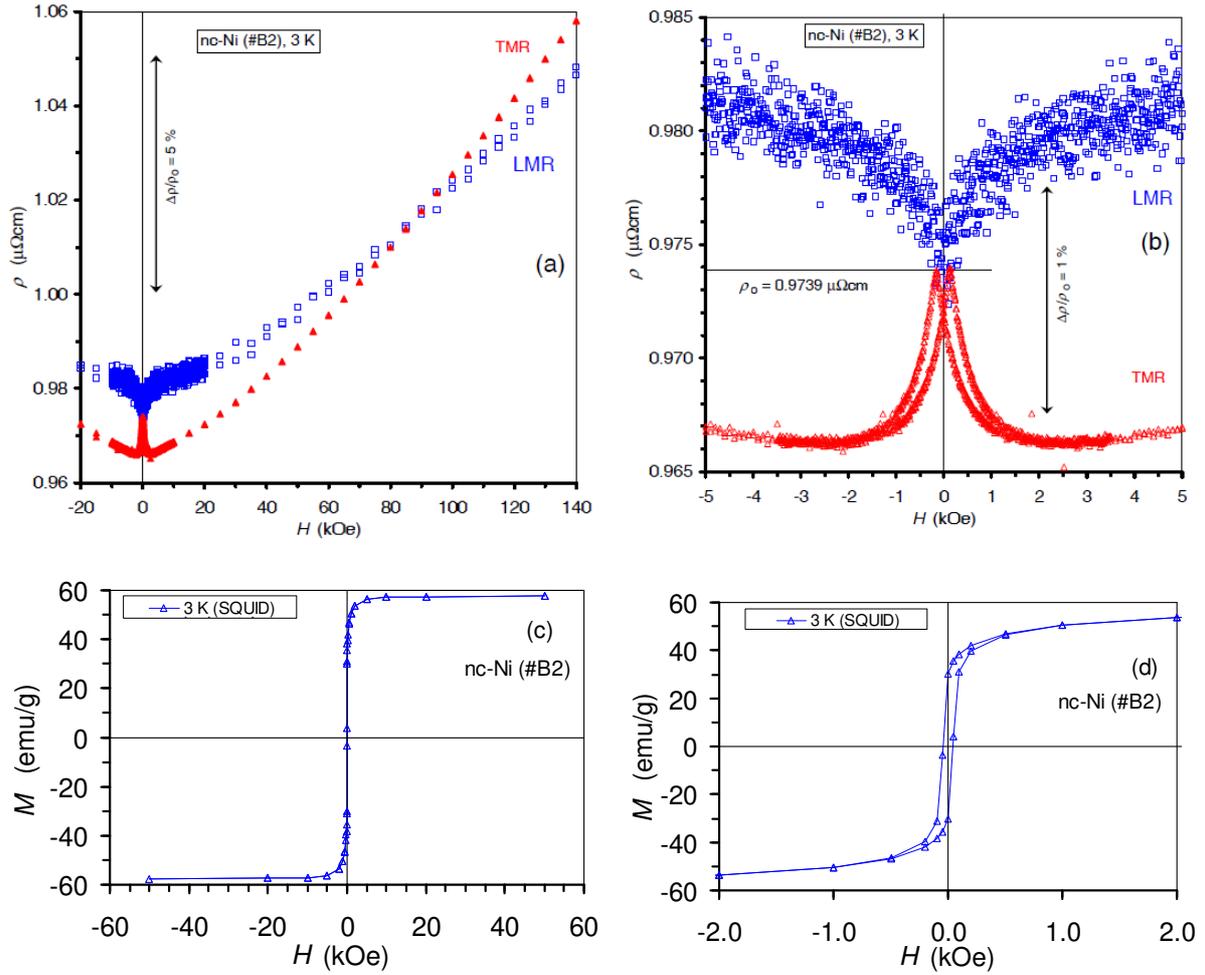

***Figure 5*** *(a) Field-dependence of the resistivity ρ at T = 3 K for nc-Ni with magnetic field orientations as indicated (LMR, TMR) in the magnetic field range from -20 kOe to +140 kOe. (b) Enlarged version of the low-field region of the MR(H) data from -5 kOe to +5 kOe. (c) High-field magnetization curves M(H) of the nc-Ni sample at 3 K. (d) Low-field magnetization curves M(H) of the nc-Ni sample at 3 K. Note: the vertical double-headed arrows in (a) and (b) indicate the specified relative resistivity change Δρ/ρ$_o$.*

Besides the large difference in the magnitude of the zero-field resistivities of the μc-Ni and nc-Ni samples at $T = 3$ K [$ρ_o$(nc-Ni) ≈ 50 $ρ_o$(μc-Ni)], one can easily recognize further important differences also in the field dependence of the resistivities for the two microstructural states of Ni when comparing Fig. 2 with Fig. 5. First, the low-field MR(*H*) behaviors are different: for μc-Ni, the resistivity has a minimum for both the LMR and TMR components around *H* = 0 and both components increase for all magnetic fields starting from *H* = 0 whereas for nc-Ni, the LMR increases and the TMR decreases at low fields just as known for the high-temperature behavior of Ni [24,39,40]. Second, at $T = 3$ K, the resistivity tends to show saturation towards high fields for both MR components for µc-Ni whereas the MR($H$) curves of nc-Ni show a permanent increase with magnetic field and their upwards curvature implies that the rate of increase becomes larger and larger with increasing magnetic field. Third, the LMR($H$) and TMR($H$) curves do not cross each other for µc-Ni whereas the rate of increase of the resistivity with magnetic field is definitely larger for TMR than for LMR for nc-Ni, so that at around $H = 90$ kOe, the TMR($H$) curves even cross the LMR($H$) data. Fourth, the relative change of the resistivity due to a magnetic field in the magnetically saturated region is much smaller for the nc-Ni sample; at $H = 70$ kOe, for example, $\Delta\rho/\rho_0 \approx 3$ % (both LMR and TMR) for nc-Ni whereas $\Delta\rho/\rho_0 \approx 50$ % (LMR) and 100 % (TMR) for µc-Ni.

All the above listed differences experienced at 3 K between µc-Ni and nc-Ni both in the magnitude of the zero-field resistivities and in the field evolutions of the resistivity are evidently a consequence of their strongly different electron mean free paths ($\lambda$). This is, on the other hand, a consequence of the difference in the microstructural states in that the nc-Ni sample has a large density of lattice defects in the form of grain boundaries whereas such defects are almost completely missing in the µc-Ni sample. It is generally considered [41] that for a given metal the product of the zero-field resistivity ($\rho_0$) and the electron mean free path is a constant for any temperatures. According to the theoretical calculations of Gall [41], for Ni metal we have $\rho_0\lambda = 4.07 \cdot 10^{-16}$ $\Omega m^2$ from which we get $\lambda$(µc-Ni;3K) = 1832 nm with our measured $\rho_0$(µc-Ni;3K) = 0.02222 µΩcm value. We get similarly for nc-Ni that $\lambda$(nc-Ni;3K) = 42 nm with our measured $\rho_0$(nc-Ni;3K) = 0.9739 µΩcm value. This means that the mean free path value for our nc-Ni sample is about half of the estimated grain size (100 nm) which may be due to some intragrain defects the presence of which is evidenced by the average crystallite size of about 75 nm deduced from XRD line broadening for this particular nc-Ni sample [23].

The much larger magnetoresistance ratio of µc-Ni than that of nc-Ni can be straightforwardly explained by the microstructural differences of the two samples. Due to the large grain size (well in the micrometer range) of the µc-Ni sample, the grain size is comparable or eventually even much larger than the electron mean free path (1832 nm) estimated for our µc-Ni sample at this temperature. As a consequence, the electrons subjected to the Lorentz force in a magnetic field are forced to turn around the magnetic field lines between the rare background scattering events (the larger the magnetic field, the more rotations per unit length along the travelling direction) and this induces a large resistivity increase. In the nc-Ni sample with a grain size of about 100 nm, the estimated electron mean free path implies that an electron is scattered about once within the grain and once at the grain boundary, travelling 42 nm between two scattering events. Due to the frequent background scattering events in nc-Ni, the rotational effect of the Lorentz field does not affect significantly the electron motion towards lower electric potentials or at least to a much lesser extent than in the µc-Ni sample and this clearly explains the much lower magnetoresistance ratio of nc-Ni in a given magnetic field as shown above.

The most striking difference between the two samples is the saturating and non-saturating character of the MR($H$) curves for µc-Ni and nc-Ni, respectively, but this cannot be explained in such a straightforward manner as we could do it for the magnitude of the magnetic-field-induced resistivity change.

Before passing on to the analysis of the MR($H$) data with help of the Kohler plot, it is mentioned that the behavior of resistivity with magnetic field in our nc-Ni sample at $T$ = 3 K is very similar to what was observed for some Ni alloys at low temperatures. In the latter cases, the increase of the zero-field residual resistivity was achieved by introducing a sufficient concentration of impurities into the pure metal matrix, resulting in a similar reduction of the electron mean free path as caused in our nc-Ni sample by the structural defects (mainly grain boundaries). To demonstrate this similarity, we will quote here available relevant literature results. In Fig. 9 of their paper, Schwerer and Silcox [31] reported the LMR($H$) and TMR($H$) data at 4.2 K for a Ni(1300ppm C) sample up to nearly $H$ = 20 kOe magnetic field which strongly resemble the corresponding MR($H$) curves in Fig. 5a for our nc-Ni sample. The agreement includes the evolution of resistivities before magnetic saturation (the immediate increase of LMR and decrease of TMR starting from $H$ = 0), the increase and upward curvature of both MR($H$) curves with clear sign of a crossing of the LMR and TMR data at higher magnetic fields. The residual resistivity of this Ni(1300ppm C) sample [31] was given as 0.449 µΩcm and this corresponds roughly to RRR =14. Both the residual resistivity and the RRR values are fairly close to the corresponding data for our nc-Ni sample (0.972 µΩcm and 9, respectively). From the residual resistivity and the extrapolated zero-induction resistivity anisotropy splitting between the LMR and TMR components, an AMR ratio of +1.8 % was deduced for the Ni(1300ppm C) sample [31]. Another classical example is the case of a $Ni_{99.42}Co_{0.58}$ alloy (the subscripts refer to atomic percentages) [4,42] for which the evolution of the LMR($H$) and TMR($H$) curves at 4.2 K as well as the residual resistivity of about 0.6 μΩcm and RRR = 13.7 are again very similar to the results on our nc-Ni sample. The resistivity anisotropy splitting can be estimated from Fig. 1c of Ref. 42 which then yields AMR = 5 %. This large value is due to the relatively large concentration of the alloying element Co which is known to cause the largest increase of the AMR ratio among all elements when added to Ni [3-5,24,40,43].

Finally, we should make a remark on the early magnetoresistance results of McGuire [42] at $T$ = 4.2 K up to $H$ = 25 kOe on a wire-shaped pure Ni polycrystalline sample with a residual resistivity ratio of 83 and residual resistivity of 0.087 μΩcm. In the magnetically saturated field range (above about 5 kOe), the relatively few data points measured indicate an approximately linear increase of the resistivity with magnetic field and with definitely higher slope for the TMR($H$) component, resulting in a crossover with the LMR($H$) data at around 10 kOe. Although the reported impurity content was fairly low (claimed to be less than 50 ppm of Mn, Co, and Fe impurities) and the wire was well annealed before the MR measurement [42], the residual resistivity of this sample is higher by a factor of 4, indicating some other residual defects or impurities, in comparison with our μc-Ni sample having a residual resistivity of 0.02222 μΩcm. This may explain that the observed MR($H$) behavior reported in Ref. 42 is different from our μc-Ni sample and rather resembles that of our nc-Ni sample, especially the higher slope of the TMR component with respect to the LMR component.

As was shown in Section 3.2 for the μc-Ni sample, if the RRR value is as high as several hundreds, an MR($H$) behavior with saturating LMR and TMR components can be observed for pure Ni. On the other hand, for sufficiently low RRR values, a behavior corresponding to our nc-Ni sample can be expected. Apparently, there is a borderline RRR value for Ni metal which separates the regimes with long or short electron mean free paths from the viewpoint of the effectiveness of the Lorentz force inducing an OMR effect. Since the Ni sample of McGuire [42] exhibiting a similar MR($H$) behavior to our nc-Ni sample has RRR = 83, we may roughly set about 100 as the borderline RRR value separating the two kinds of MR($H$) behavior.

It is noted furthermore in connection with the Ni sample of McGuire [42] that the absence of both a saturating character and an upward curvature of the MR($H$) curves, but rather a nearly linear field variation of the resistivity also suggest that this sample is indeed in the vicinity of the borderline RRR value from the magnetoresistance point of view. McGuire [42] reported an AMR ratio of +2.2 % which is close to our result, but we cannot consider it as a reliable value for the low-temperature AMR ratio of pure bulk Ni due to the way of determination. Namely, for calculating the AMR ratio, McGuire [42] took the difference between the $\rho_L$ and $\rho_T$ values arbitrarily at $H$ = 4 kOe which is incorrect. This is because whereas due to the wire-shaped specimen, the $\rho(B=0)$ condition is approximately fulfilled for the TMR component, an extrapolation to zero induction would have been necessary for the LMR component..

The Kohler plot for the LMR and TMR data at $T$ = 3 K for the nc-Ni sample is shown in Fig. 6. This plot also reveals a stronger increase of the resistivity with magnetic field for the TMR component than for the LMR one. A similar fitting analysis was carried out for the data of nc-Ni as was done for the μc-Ni sample in the previous section with the difference that, due to the about 50 times larger residual resistivity of the nc-Ni sample, the resistivity data are now less scattered and a third-order polynomial fitting was sufficient since it provided even better fit quality parameter values ($R^2$) than the fourth-order polynomial for the μc-Ni sample.

Furthermore, due to the larger residual resistivity, the extrapolation to $B$ = 0 on the $B/\rho(B=0)$ axis could be carried out from much lower abscissa values and all these improvements resulted in more accurate values for $\rho(B=0)$ of both the LMR and TMR components from which we can get more reliable values for $\Delta\rho_{AMR}$ and for the AMR ratio. All these derived data are included in the textboxes in Fig. 6.

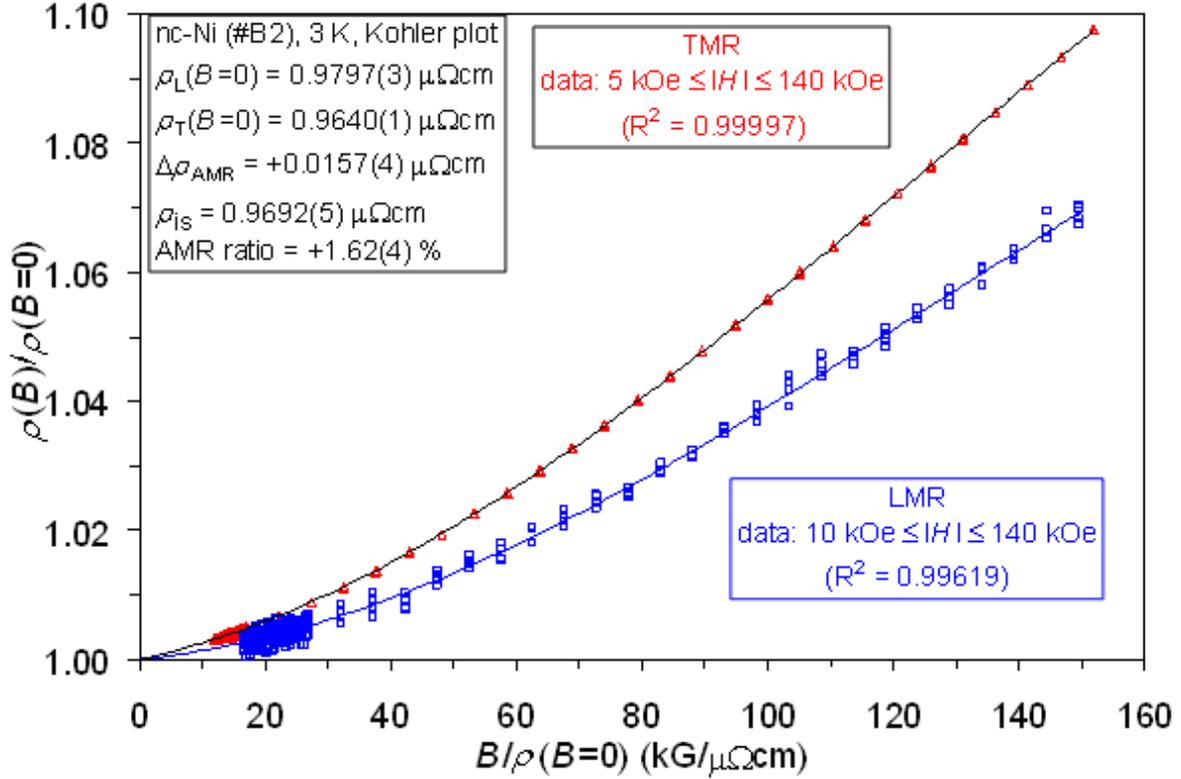

***Figure 6***  Kohler plot $\rho(B)/\rho(B=0)$ vs. $B/\rho(B=0)$ at $T = 3$ K for nc-Ni with magnetic field orientations as indicated (LMR, TMR) for the MR(H) data in the magnetic field range  10 kOe $\leq |H| \leq$ 140 kOe (LMR) and 5 kOe $\leq |H| \leq$ 140 kOe (TMR). The experimental data are the symbols (squares: LMR; triangles: TMR), the solid lines are the third-order polynomial fitting functions providing an empirical analytical form of the Kohler function F. The fitting parameter values were obtained at the maximum values of the normalized fit quality parameter $R^2$ specified in the text boxes attached to the LMR and TMR curves. The extrapolated zero-induction resistivities $\rho(B=0)$ for the LMR and TMR components, the isotropic resistivity $\rho_{is}$ as well as the derived resistivity anisotropy parameters ($\Delta\rho_{AMR}$ and AMR ratio) are given in the text box in the upper left corner (see text for more details).

We have prepared the $\rho$ vs. $B$ plot also for the nc-Ni sample (Fig. 7). Due to the significantly higher accuracy of the MR(*H*) data for the nc-Ni sample, as a consequence of its much higher residual resistivity, when leaving the $\rho(B=0)$ parameter as a free variable for the fitting, we got the same $\rho(B=0)$, $\Delta\rho_{AMR}$ and AMR values (within about 0.1 %) as from the Kohler plot analysis which again justifies the reliability of the latter data.

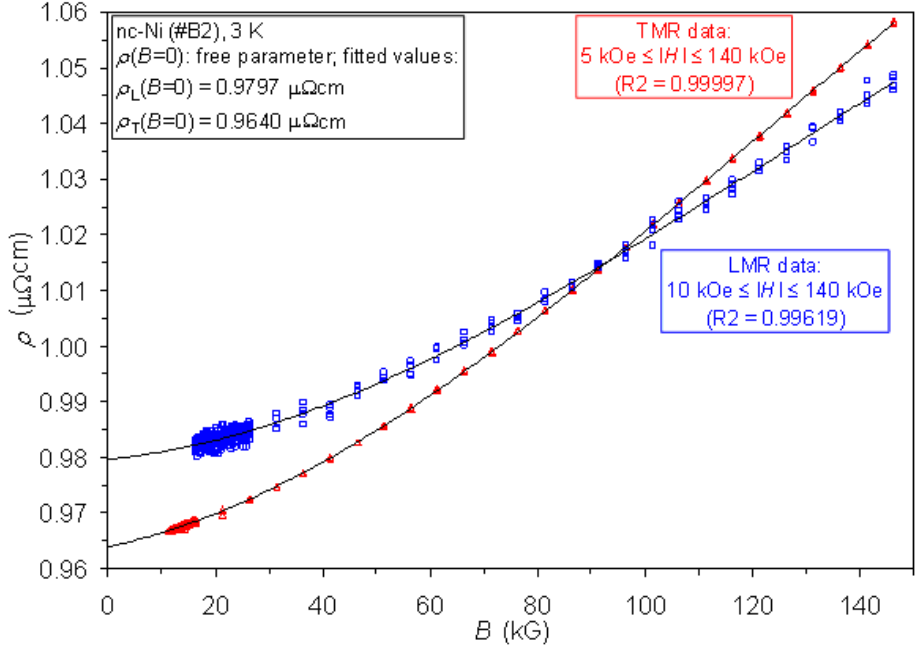

***Figure 7***  *Resistivity $\rho$ vs. B at T = 3 K for nc-Ni with magnetic field orientations as indicated (LMR, TMR) for the MR(H) data in the magnetic field range 10 kOe $\leq |H| \leq$ 70 kOe (LMR) and 5 kOe $\leq |H| \leq$ 70 kOe (TMR). The experimental data are the symbols (squares: LMR; triangles: TMR), the solid lines are the third-order polynomial fitting functions providing an accurate description of the field evolution of the measured data. During fitting, the zero-induction resistivities $\rho(B=0)$ for the LMR and TMR components were free variables and the same values were obtained as derived from the Kohler plot (Fig. 6).*

The $\rho$ vs. *B* plot in Fig. 7 clearly demonstrates that $\rho_L(B=0) > \rho_T(B=0)$, i.e., the nc-Ni sample has definitely a positive AMR. Furthermore, when displaying the MR(*H*) data on a $\rho$ vs. *H* plot and letting $\rho(H=0)$ as a free parameter, we get $\rho_L(H=0)$ and $\rho_T(H=0)$ values which yield an AMR ratio of +1.54 % which value is quite close to +1.62 %, the value deduced from the Kohler plot by extrapolating the resistivity to *B* = 0 (cf. Fig. 6). This implies that the OMR contribution to the field dependence of the resistivity for the present nc-Ni sample even at a temperature as low as *T* = 3 K is fairly small, although it is still the only source of a magnetic-field-induced resistivity change in the monodomain state since at this temperature, there is no magnetic disorder. This is in very good agreement with the considerations put forward earlier in this section when comparing the different microstructural features and the concomitant mean free path difference between the µc-Ni and nc-Ni samples investigated in the present work.

## 3.4 Comparison of the magnetic-field-induced resistivity change in µc-Ni and nc-Ni at T = 3 K and comparison with relevant reported experimental data

A comparison of Figs. 2a and 5a a reveals that the magnetic-field-induced resistivity changes at 3 K for the µc-Ni and nc-Ni samples are of comparable magnitude although the field dependencies of the resistivity for the two samples are quite different. As demonstrated in Figs. 4 and 7, the variation of the resistivity with magnetic induction could be properly described for the measured data in the magnetically saturated (monodomain) state by an empirical fourth-order polynomial fitting function $\rho(B) = \rho(B=0) + \alpha \cdot B + \beta \cdot B^2 + \gamma \cdot B^3 + \delta B^4$ for the µc-Ni sample whereas a third-order polynomial fitting function (i.e., $\delta = 0$) was sufficient for the nc-Ni sample. For convenience, the values of the parameters $\alpha$, $\beta$, $\gamma$ and $\delta$ as well as the $\rho(B=0)$ values are collected in Table I.

*Table I* *Parameters characterizing the field dependence of the induced resistivity change $\Delta\rho(B) = \rho(B) - \rho(B=0)$ of the µc-Ni and nc-Ni samples at 3 K as derived in Figs. 4 and 7. The parameter values were obtained as explained in the text from fits of the experimental resistivity data in the magnetically saturated region for µc-Ni and nc-Ni to the empirical function $\rho(B) = \rho(B=0) + \alpha \cdot B + \beta \cdot B^2 + \gamma \cdot B^3 + \delta B^4$ by fixing the corresponding $\rho(B=0)$ values as determined from the Kohler plot analysis.*

| T = 3 K fitted parameters | | $\rho(B=0)$ (µΩ·cm) | $\alpha$ (µΩ·cm/kG) | $\beta$ (µΩ·cm/(kG)$^2$) | $\gamma$ (µΩ·cm/(kG)$^3$) | $\delta$ (µΩ·cm/(kG)$^4$) |
|---|---|---|---|---|---|---|
| µc-Ni (#B5) | LMR | 0.01918 | +5.875·10$^{-4}$ | -1.350·10$^{-5}$ | +1.668·10$^{-7}$ | -8.094·10$^{-10}$ |
| | TMR | 0.01896 | +8.358·10$^{-4}$ | -1.434·10$^{-5}$ | +1.664·10$^{-7}$ | -7.901·10$^{-10}$ |
| nc-Ni (#B2) | LMR | 0.9797 | +0.897·10$^{-4}$ | +0.415·10$^{-5}$ | -0.110·10$^{-7}$ | 0 |
| | TMR | 0.9640 | +2.012·10$^{-4}$ | +0.501·10$^{-5}$ | -0.137·10$^{-7}$ | 0 |

In order to better visualize the different behaviors of the field dependence of the resistivity for the two microstructural states of Ni, the induced resistivity change $\Delta\rho(B) = \rho(B) - \rho(B=0)$ is displayed in Fig. 8a as a function of the magnetic induction $B$ for both samples. This means that for each component (LMR and TMR), the resistivity change $\Delta\rho(B)$ is referred to its zero-induction value.

Both samples exhibit a resistivity increase of comparable magnitude up to not very high magnetic inductions, but since the curvature is quite different for the two samples at higher inductions, the divergence of the datasets for the two different microstructural states is strongly enhanced (Fig. 8a).

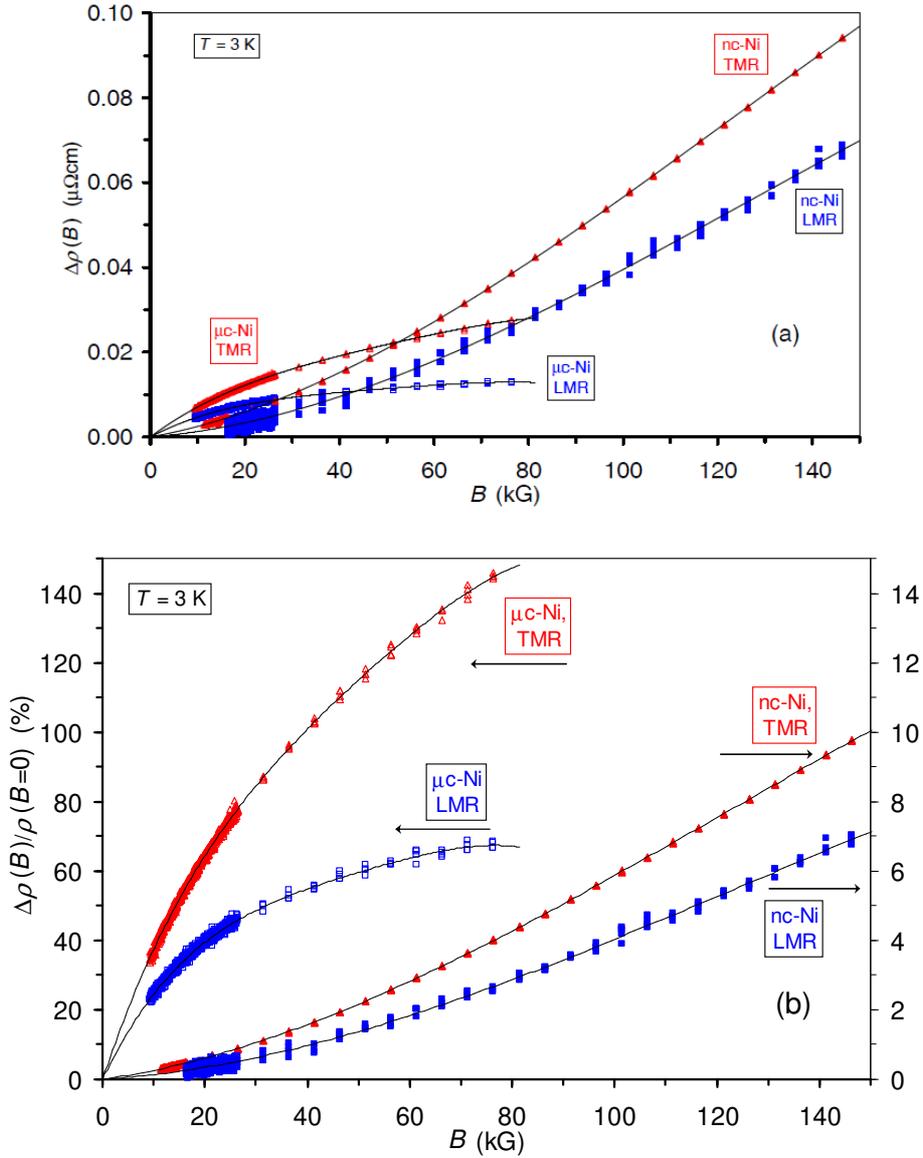

***Figure 8*** *(a) Resistivity change $\Delta\rho(B) = \rho(B) - \rho(B=0)$ vs. B in the magnetically saturated state at T = 3 K for µc-Ni and nc-Ni with magnetic field orientations as indicated (LMR, TMR). The experimental data are the symbols (squares: LMR; triangles: TMR), the solid lines are the corresponding fitting functions with the same of polynomial order as in the Kohler plot. (b) The same data displayed as the relative change of the resistivity in the form of $\Delta\rho(B)/\rho(B=0)$ vs. B plots. The reference resistivity in each case was the zero-induction resistivity $\rho(B=0)$ determined for the given measurement configuration (LMR and TMR) from a Kohler plot analysis. Note the different ordinate scales for the µc-Ni (left axis) and nc-Ni (right axis) samples in (b).*

As discussed already earlier in this section, the different behaviors of the resistivity variation with magnetic induction for the two samples reflect the differences in their electron mean free paths what, on the other hand, is a direct consequence of their different microstructural states. Furthermore, we can also see that for any induction value $B$, the relation $\Delta\rho$(TMR) > $\Delta\rho$(LMR) holds for both samples in the magnetically saturated (monodomain) state as expected for a dominant OMR contribution to the field-induced resistivity change.

Figure 8b shows the relative change $\Delta\rho(B)/\rho(B=0)$ with magnetic induction for the two samples. With reference to the different left-hand and right-hand scales, it can be clearly seen that at such a low temperature the relative resistivity change is much larger for the μc-Ni sample having low resistivity than for the nc-Ni sample which has a substantial residual resistivity.

According to Fig. 9, the shape of the Kohler plots and the relation of the LMR and TMR datasets are the same for our μc-Ni sample and for the bulk Ni series of Ref. 30. The quantitative differences between the results for a given magnetic field orientation (LMR or TMR) are due to the different kind and concentration of impurities in our μc-Ni sample and in those of Schwerer and Silcox [30]. Unfortunately, the zero-induction resistivities $\rho_L(B=0)$ and $\rho_T(B=0)$ were not reported in the latter paper, so no AMR ratio can be assessed from these data to compare with our result.

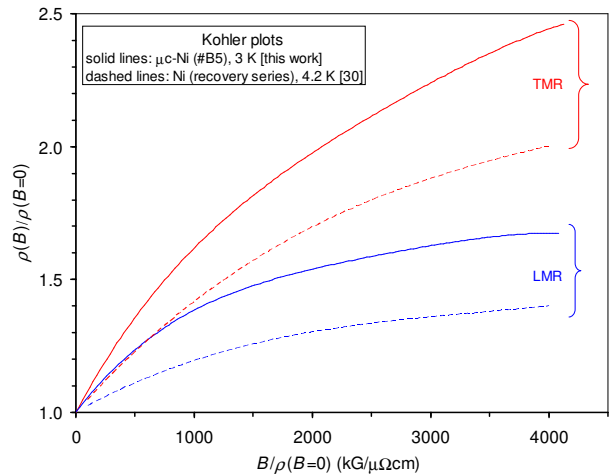

***Figure 9*** *Comparison of the Kohler plots $\rho(B)/\rho(B=0)$ vs. $B/\rho(B=0)$ at low temperatures for the present μc-Ni sample with the data of Schwerer and Silcox [30] on their "recovery series" samples (see text for more details).*

As to the nanocrystalline state, the only available relevant report on a nc-Ni sample is the work of Madduri and Kaul [22] who observed an increase of the transverse component of the magnetoresistance with magnetic field at $T = 2$ K for nc-Ni samples with crystallite sizes from 10 to 40 nm similarly as in our study at $T = 3$ K. However, the measured resistivity increase at their maximum field of $H = 90$ kOe was by about a factor of 250 smaller for the sample with 40 nm crystallite size than the corresponding value on our nc-Ni sample with a crystallite size of 75 nm although the residual resistivities in zero magnetic field were quite comparable for the two samples (our nc-Ni sample: 0.9739 μΩcm; nc-Ni sample in Ref. 22: 0.42 μΩcm). We have no explanation for the discrepancy in these low-temperature results.

It is worth mentioning, however, that Rüdiger et al. [44] reported longitudinal and transverse Kohler plots at $T = 1.5$ K for a 2-μm-wide Fe wire lithographically patterned from a 100-nm-thick bcc-Fe film epitaxially grown on a sapphire substrate which are very similar to our Kohler plots for the nc-Ni sample (Fig. 6). The zero-field resistivity of the Fe wire was 0.74 μΩcm at $T = 1.5$ K (yielding RRR = 20) which compares well with our 0.9739 μΩcm value for the nc-Ni sample (RRR = 9). The similar resistivity and RRR values of the pure Fe wire indicate that it may consist of fine grains and/or surface scattering effects may also yield a finite contribution to the resistivity which altogether yield a reduced electron mean free path with respect to pure bulk Fe. Therefore, the results of Rüdiger et al. [44] on a nanoscale Fe wire give support to the findings on our nc-Ni sample in that the behavior reported in the present work is indeed a general phenomenon. This means that, in conformity with the results of early studies on some Ni alloys as discussed in subsection 3.3 (see, e.g., the results on the Ni(1300ppm C) sample of Ref. 31), a very similar MR behavior can also be achieved in a pure metal by reducing the electron mean free path via the introduction of excess static scatterers in the form of lattice defects (e.g., grain boundaries) and/or via finite size effects (electron scattering at external surfaces).

## 4. Electrical transport and magnetic properties at $T = 300$ K

*4.1 Magnetoresistance and magnetic properties of µc-Ni at T =300 K*

Figure 10a shows the field dependence of the resistivity for µc-Ni at 300 K for the LMR and TMR measurement configurations as measured up to $H = 70$ kOe, in qualitative agreement with the well-known room-temperature MR($H$) curves reported for bulk Ni up to about $H = 18$ kOe [39]. The low-field sections of the measured magnetoresistance curves are shown in Fig. 10b in the form of magnetoresistance ratio $\Delta\rho/\rho_o$ vs. magnetic field. As for the low-temperature measurements, the resistivity $\rho_o$ is identified with the peak value of the resistivity, i.e., $\rho_o = \rho(H_p)$, for MR($H$) curves with a clear hysteresis. It should be noted, nevertheless, that due to the high accuracy of the measurement setup, the resistivity peak values measured for the LMR and TMR components agreed within about 0.2 %.

The low-field MR($H$) data in Fig. 10b reflect the magnetization process [3,4] the field evolution of which depends on sample shape, magnetic anisotropy, stresses, etc. To illustrate this, Fig. 10c displays the room-temperature magnetization curve $M(H)$ of the µc-Ni foil up to $H = 50$ kOe as measured by SQUID which shows that technical saturation is achieved around a few kilooersted of magnetic field. In compliance with this, the break in the MR($H$) curves roughly at around $H = 1$ kOe magnetic field (Fig. 10b) indicates that saturation is achieved for both the LMR and TMR components at similarly small magnetic fields as was found for the magnetization. It can be seen furthermore from Figs. 10a and 10b that in the saturated (monodomain) state, the field dependence of the MR($H$) curves is very similar for the LMR and TMR components.

From the MR($H$) curves in Fig. 10b, the peak positions were found to be at $H_p = \pm 75$ Oe for the LMR component at room temperature (the TMR curves were too broad to determine the exact peak positions).

The SQUID magnetometer could not reveal the hysteresis either at 300 K due to the low coercive field of the µc-Ni sample. Therefore, Fig. 10d shows the low-field section of the $M(H)$ curve measured by VSM at 300 K which reveals a hysteresis with a coercive field of $H_c = 13$ Oe.

The experimental results for the magnetoresistance of µc-Ni at $T = 300$ K correspond well to the schemes depicted in Fig. 3 of Ref. 25 for the high-temperature behavior of ferromagnets (at room temperature, for nickel metal with $T_C = 631$ K, we have $T = 300$ K ~

0.5 $T_C$). One can also observe at $T = 300$ K an approximately linear decrease of the resistivity above the saturation field.

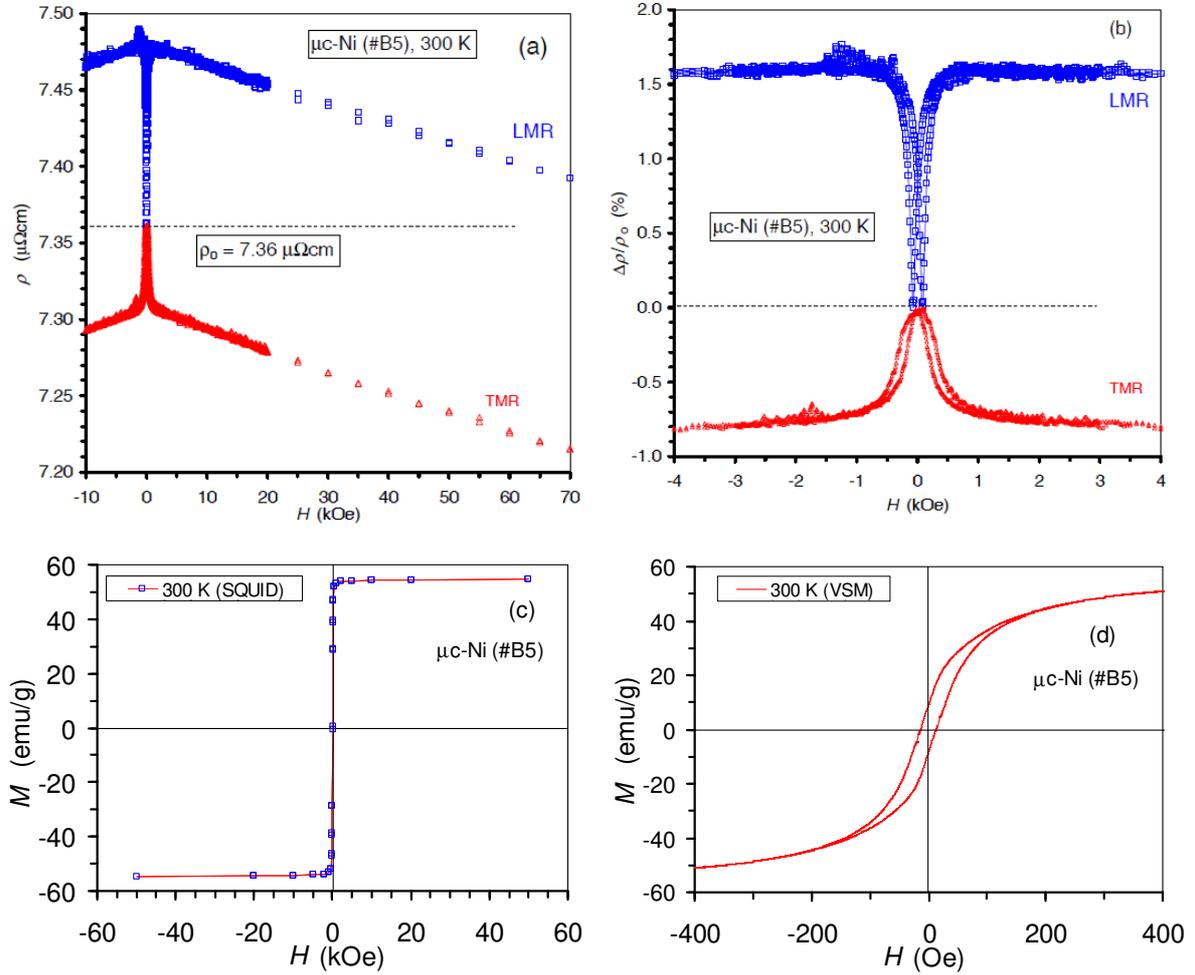

*Figure 10* *(a) Field-dependence of the resistivity $\rho$ at $T = 300$ K for the µc-Ni sample with magnetic field orientations as indicated (LMR, TMR) in the magnetic field range from -10 kOe to +70 kOe; (b) Low-field resistivity data in the form of magnetoresistance ratio $\Delta\rho/\rho_o$ vs. magnetic field where $\rho_o = \rho(H_p)$; c) High-field magnetization curve M(H) at 300 K for the µc-Ni sample; (d) Low-field magnetization curve M(H) at 300 K for the µc-Ni sample.*

The decrease of the resistivity with increasing magnetic field in the saturation region is due to the gradual suppression of the thermally-induced spin disorder [22,33,34] since at finite temperatures the scattering of conduction electrons on non-aligned individual magnetic moments also gives a contribution to the resistivity. By increasing the magnetic field after technical saturation (in the monodomain state), the thermally disordered magnetic moments are more and more aligned along the magnetic field [4] (this is often termed also as paraprocess [33]) and, therefore, this kind of scattering is diminished and, thus, one can observe a resistivity decrease.

We have displayed the room-temperature MR($H$) data of the µc-Ni sample for $|H| \geq 3$ kOe on a Kohler plot in Fig. 11. A second-order polynomial was found to be satisfactory for a sufficiently accurate fitting of the data. All the fitted and derived parameters are specified in Fig. 11. It is noted that the derived parameters agree fairly well with the room-temperature data reported in our previous work [24] for the same µc-Ni sample: $\Delta\rho_{AMR}$ = +0.176(12) µΩcm, $\rho_0$ = 7.36(21) µΩcm and AMR = +2.39(16) %. In that work, a detailed comparison with previous bulk Ni data was also discussed.

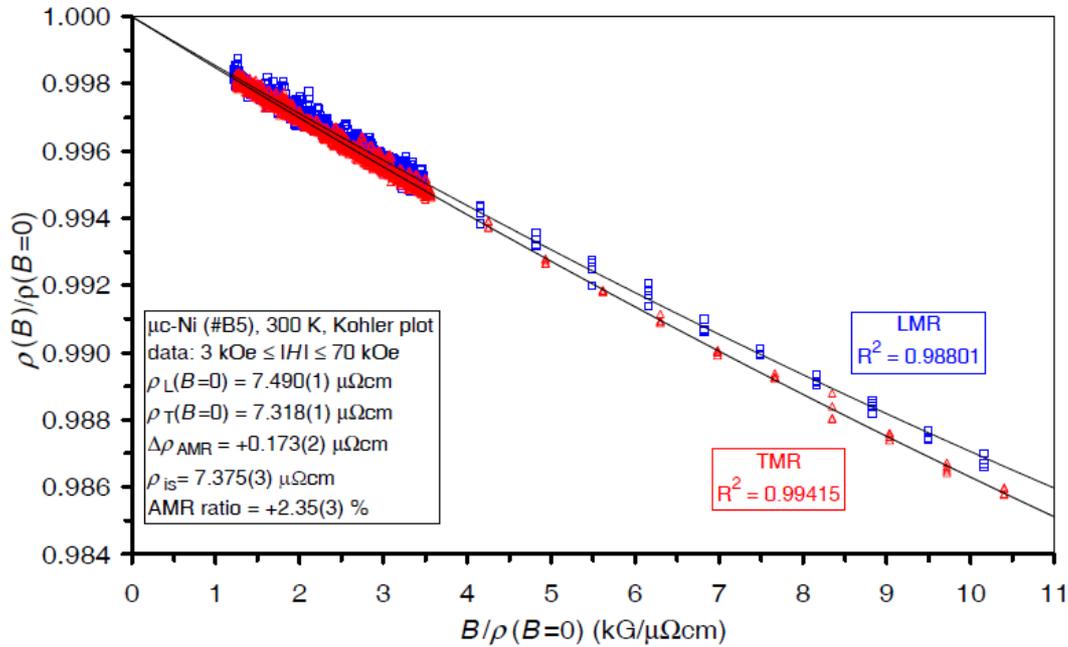

*Figure 11 Kohler plot $\rho(B)/\rho(B=0)$ vs. $B/\rho(B=0)$ at T = 300 K for µc-Ni with magnetic field orientations as indicated (LMR, TMR) for the MR(H) data in the magnetic field range 3 kOe ≤ |H| ≤ 70 kOe. The experimental data are the symbols (squares: LMR; triangles: TMR), the solid lines are the second-order polynomial fitting functions providing an empirical analytical form of the Kohler function F. The fitting parameter values were obtained at the maximum values of the normalized fit quality parameter $R^2$ specified in the text boxes attached to the LMR and TMR curves. The extrapolated zero-induction resistivities $\rho(B=0)$ for the LMR and TMR components, the isotropic resistivity $\rho_{is}$ as well as the derived resistivity anisotropy parameters ($\Delta\rho_{AMR}$ and AMR ratio) are given in the text box in the lower left corner (see text for more details).*

We have displayed the same data for the magnetic field range $|H| \geq 3$ kOe also on a $\rho$ vs. $B$ plot (Fig. 12) which properly reveals the resistivity anisotropy splitting between the LMR and TMR components in the magnetically saturated state. Similarly to the Kohler plot (Fig. 11), a second-order polynomial fit provided a sufficiently accurate description of the experimental data. The $\rho(B=0)$ values were allowed as free parameters and the same fitted values were obtained as derived from the Kohler plot, justifying the latter data. Furthermore, a $\rho$ vs. $H$ plot was also prepared and a similar fit with $\rho(H=0)$ as free parameter yielded $\rho_{LMR}(H=0)$ and $\rho_{TMR}(H=0)$ values which agreed within 0.1 % with the corresponding $\rho_{LMR}(B=0)$ and $\rho_{TMR}(B=0)$ data, the slight difference coming from the shift of the resistivity data along the abscissa ($B = H + 4\pi M_s$ where $4\pi M_s(300K) = 6.1$ kG for Ni). The coefficients of the second-order terms agreed up to the fourth digit and the coefficients of the first-order terms agreed within 2%.

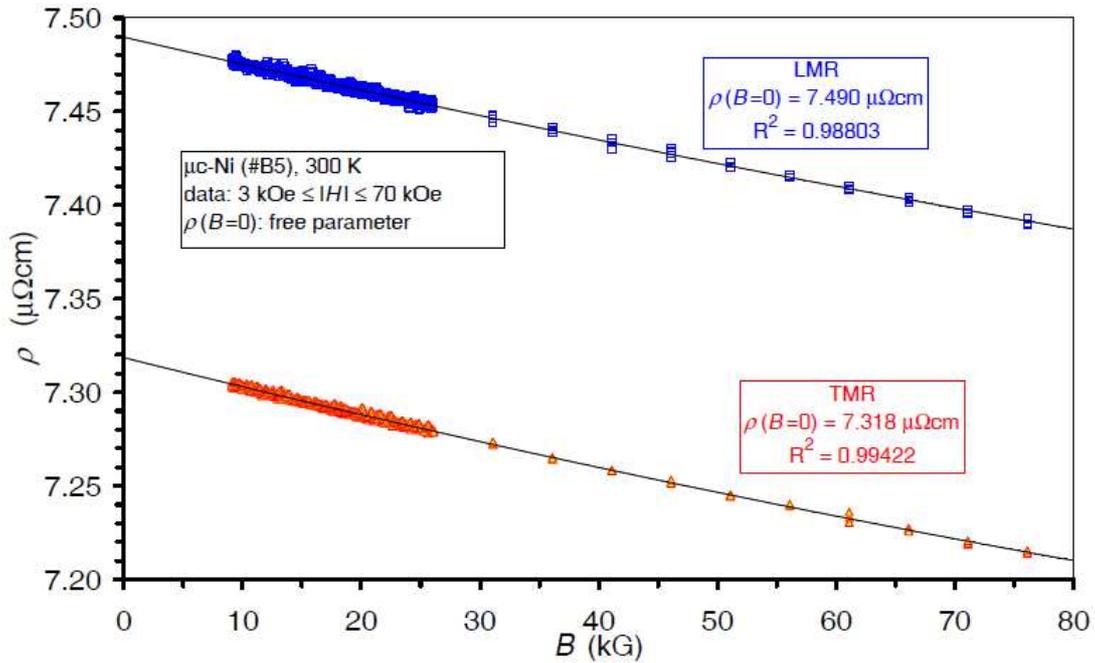

*Figure 12* *Resistivity $\rho$ vs. B at T = 300 K for $\mu$c-Ni with magnetic field orientations as indicated (LMR, TMR) for the MR(H) data in the magnetic field range 3 kOe $\leq |H| \leq$ 70 kOe. The experimental data are the symbols (squares: LMR; triangles: TMR), the solid lines are the fitting functions providing an accurate description of the field evolution of the measured data. During fitting, the zero-induction resistivities $\rho(B=0)$ for the LMR and TMR components were free variables and the same values were obtained as derived from the Kohler plot (Fig. 11).*

Since in our previous work [24] the magnetoresistance study of the same µc-Ni sample was carried out in magnetic fields up to 9 kOe only, we have analyzed the present data also in the field range from 3 kOe to 10 kOe and the AMR ratio remained the same up to the specified three digits.

The small difference between the zero-field and zero-induction resistivity data implies that in µc-Ni at $T$ = 300 K, the OMR contribution to the magnetoresistance is negligible as was suggested also from theoretical considerations [34]. The reason for this is that at room temperature, due to the strong electron-phonon scattering, the mean free path of bulk Ni metal is short. According to the theoretical calculations of Gall [41], for Ni metal we have $\rho_0 \lambda$ = $4.07 \cdot 10^{-16}$ $\Omega m^2$ from which we get $\lambda$(µc-Ni;300K) = 5.5 nm with our measured $\rho_0$(µc-Ni;300K) = 7.36 µΩcm value. It is noted that by using a free-electron estimate formula by Ashcroft and Mermin for the electron mean-free path [45], we have recently derived [46] $\lambda$(Ni;300K) = 5.71 nm which is fairly close to the above obtained value for $\lambda$(µc-Ni;300K).

### *4.2 Magnetoresistance and magnetic properties of nc-Ni at T =300 K*

Figures 13a and 13b show the field dependence of the resistivity for nc-Ni at 300 K for both measurement configurations (LMR, TMR). The experimental results for the room-temperature magnetoresistance of nc-Ni are very similar to those obtained for µc-Ni (cf. Fig. 10) and correspond well to the schemes depicted in Fig. 3 of Ref. 30 for the high-temperature MR behavior because for nickel metal ($T_C$ = 631 K) at room temperature we have $T$ = 300 K ~ 0.5 $T_C$. It is noted that the resistivity peak values measured for the LMR and TMR components agreed within about 0.06 %.

The good quality of the MR($H$) data enabled the determination of the field positions of the resistivity minima and maxima for both the LMR and TMR components (Fig. 13b) and $H_p$ = ±160 Oe was obtained for the nc-Ni sample. This is about twice as much as the value obtained for the µc-Ni sample ($H_p$ = 75 Oe) and qualitatively well matches the factor of three difference between the measured coercive fields of the two samples.

The high-field $M(H)$ curve of the nc-Ni sample as obtained by SQUID at $T$ = 300 K is displayed in Fig. 13c. The low-field sections of the $M(H)$ curves as obtained by both SQUID and VSM are displayed in Fig. 13d which reveal the same hysteretic behavior as can be seen in the MR($H$) curves (Fig. 13b). From the VSM data in Fig. 13d, $H_c$ = 37 Oe was obtained for the nc-Ni sample.

We can observe in Fig. 13b that the magnetoresistance reaches saturation along both the L and T orientations in fairly low, nearly identical magnetic fields, in compliance with the $M(H)$ curves. It was found that the saturation of the MR($H$) curves can be considered as complete for magnetic fields $|H| \geq 3$ kOe (the behavior of the MR($H$) data in this field range for nc-Ni was very similar to the behavior of µc-Ni shown in Fig. 10). Therefore, these high-field data were used for the further analysis of the field dependence of the resistivity in the magnetically saturated (monodomain) state.

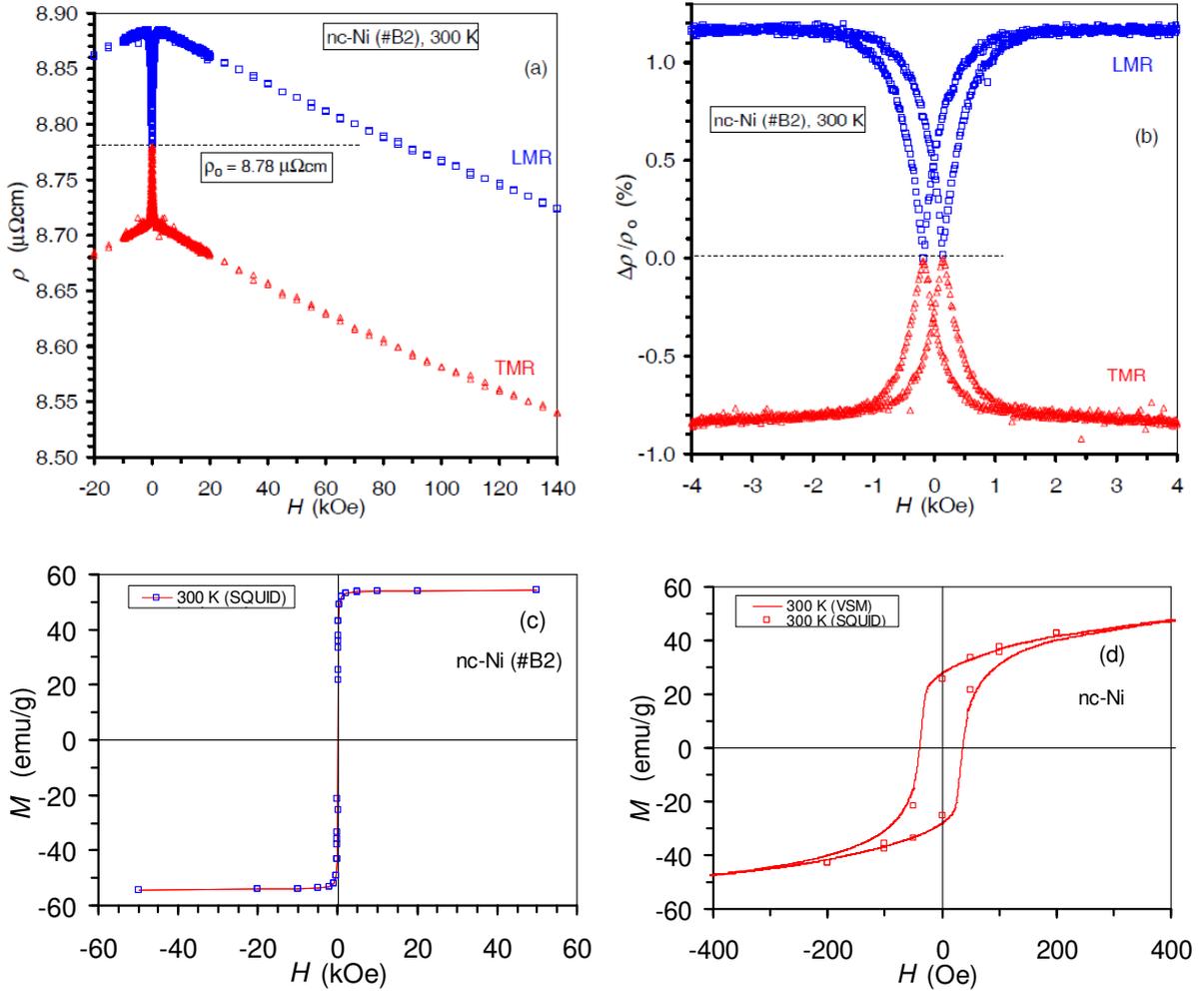

*Figure 13* (a) Field-dependence of the resistivity $\rho$ at $T = 300$ K for nc-Ni with magnetic field orientations as indicated (LMR, TMR) in the magnetic field range from -20 kOe to +140 kOe; (b) Low-field resistivity data in the form of magnetoresistance ratio $\Delta\rho/\rho_o$ vs. magnetic field where $\rho_o = \rho(H_p)$; (c) High-field magnetization curve M(H) at 300 K for the nc-Ni sample; (d) Low-field magnetization curve M(H) at 300 K for the nc-Ni sample.

Due to the qualitative similarity of the room-temperature magnetoresistance data for µc-Ni and nc-Ni, we applied exactly the same procedure for nc-Ni as used for µc-Ni in Section 4.1 when constructing and analyzing the Kohler plot and the $\rho$ vs. $B$ plot. The results are shown in Figs. 14 and 15, respectively, where all the fit results and the derived transport parameters are also displayed.

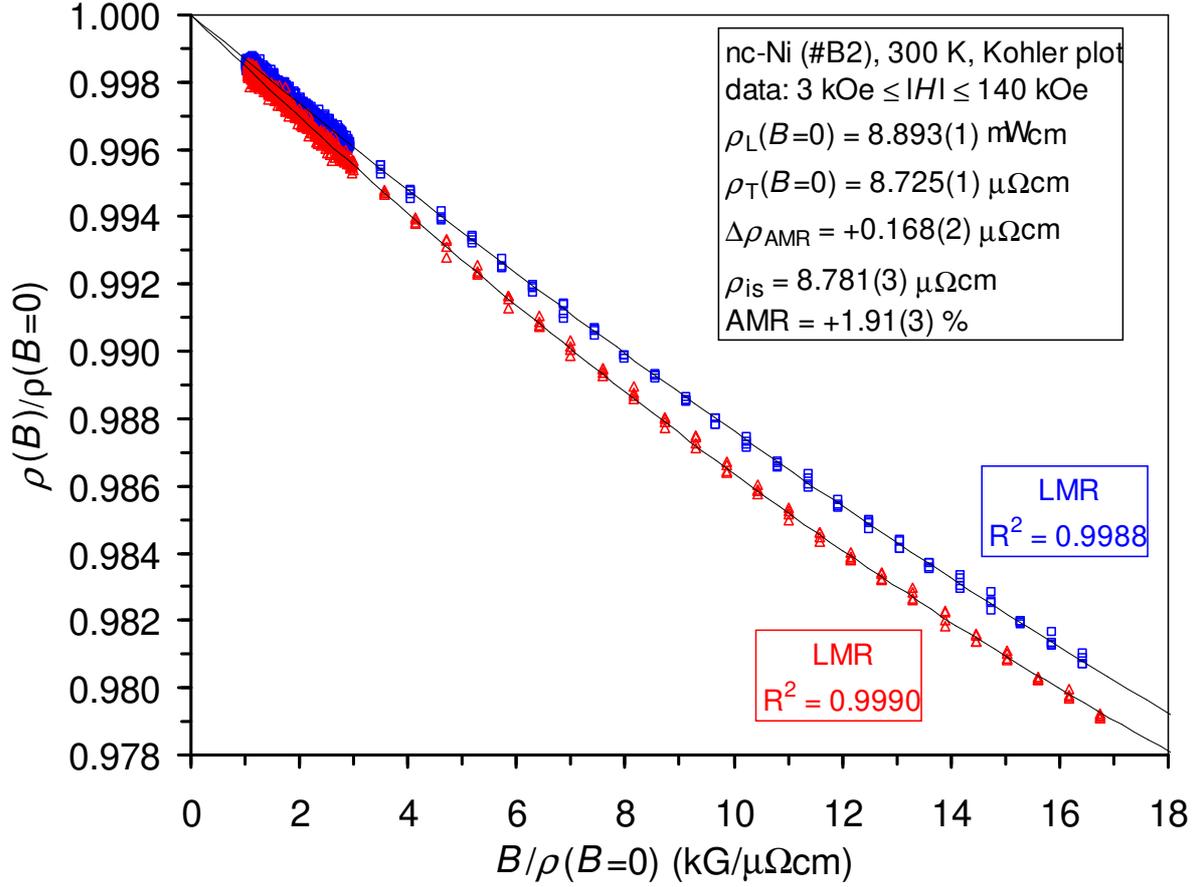

*Figure 14 Kohler plot $\rho(B)/\rho(B=0)$ vs. $B/\rho(B=0)$ at $T = 300$ K for nc-Ni with magnetic field orientations as indicated (LMR, TMR) for the MR(H) data in the magnetic field range 3 kOe $\leq |H| \leq 140$ kOe. The experimental data are the symbols (squares: LMR; triangles: TMR), the solid lines are the second-order polynomial fitting functions providing an empirical analytical form of the Kohler function F. The fitting parameter values were obtained at the maximum values of the normalized fit quality parameter $R^2$ specified in the text boxes attached to the LMR and TMR curves. The extrapolated zero-induction resistivities $\rho(B=0)$ for the LMR and TMR components, the isotropic resistivity $\rho_{is}$ as well as the derived resistivity anisotropy parameters ($\Delta\rho_{AMR}$ and AMR ratio) are given in the text box in the upper right corner (see text for more details).*

In the case of the $\rho$ vs. $B$ plot (Fig. 15), the $\rho(B=0)$ values were free parameters and the same fitted values were obtained as from the Kohler plot. Furthermore, similarly to the analysis of the MR($H$) data of µc-Ni at 300 K, when constructing the $\rho$ vs. $H$ plot for nc-Ni, there was a good agreement (within 0.1 %) of the fitted $\rho(H=0)$ values with the corresponding $\rho(B=0)$ values (with the same agreement between the coefficients of the first-order and second-order terms as for µc-Ni), implying again a negligible contribution of the OMR effect to the observed magnetoresistance at room temperature. It is noted finally that the derived parameters agree fairly well with the room-temperature data reported in our previous work [24] for the same nc-Ni sample (#B2): $\Delta\rho_{AMR}$ = +0.165 µΩcm, $\rho_0$ = 8.78 µΩcm and AMR = +1.88 %.

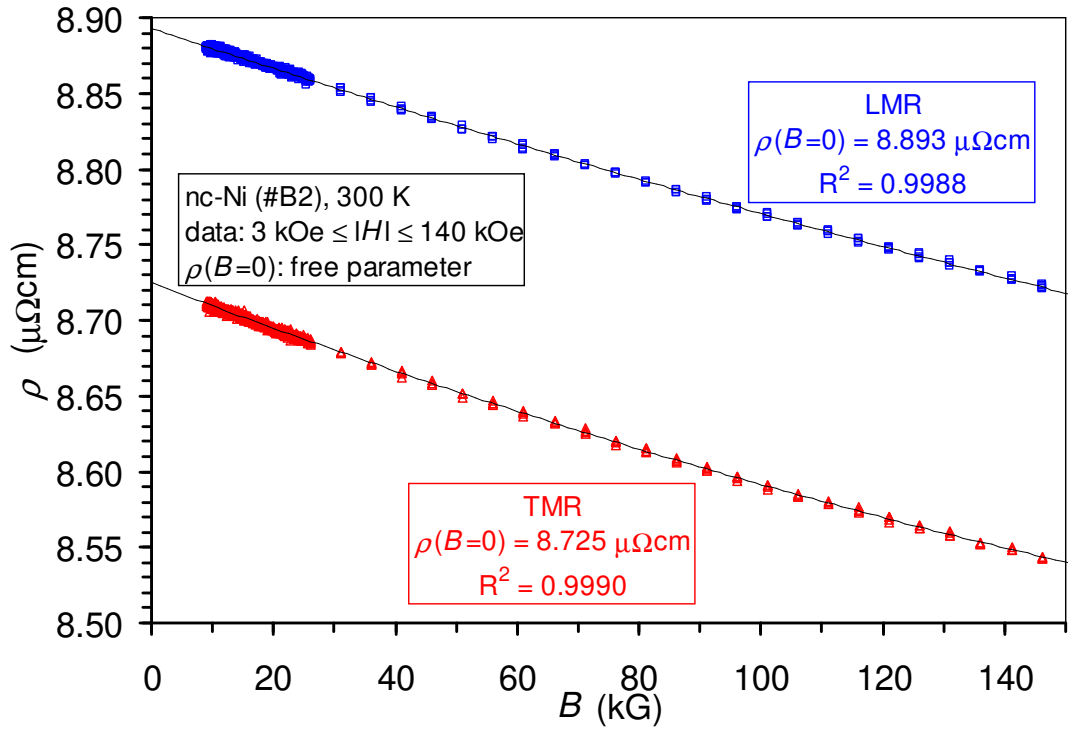

*Figure 15 Resistivity $\rho$ vs. B at T = 300 K for nc-Ni with magnetic field orientations as indicated (LMR, TMR) for the MR(H) data in the magnetic field range 3 kOe ≤ |H| ≤ 70 kOe. The experimental data are the symbols (squares: LMR; triangles: TMR), the solid lines are the second-order polynomial fitting functions providing an accurate description of the field evolution of the measured data. During fitting, the zero-induction resistivities $\rho(B=0)$ for the LMR and TMR components were free variables and the same values were obtained as derived from the Kohler plot (Fig. 14).*

*4.3 Overview of the magnetic properties of µc-Ni and nc-Ni at T = 3 K and 300 K*

It is noted first that from the magnetizations measured at 50 kOe at both $T = 3$ K and 300 K for the µc-Ni and nc-Ni samples, the saturation magnetization values were found to agree within about 0.5 % for the two structural modifications of Ni. This is in good agreement with previous reports [47-49] according to which the saturation magnetization of sufficiently pure massive Ni samples hardly changes with respect to the bulk value even in nc-Ni with grain sizes down to as low as 10 nm. In line with this, theoretical considerations [50] have also justified the negligible effect of the disorder in the grain boundaries on the saturation magnetization of Ni for such grain sizes.

Next, the hysteresis parameters of the *M(H)* and MR(*H*) curves at $T = 3$ K and 300 K as deduced in previous sections for our µc-Ni and the nc-Ni samples are summarized in Table II.

*Table II  Parameters characterizing the remagnetization process of the µc-Ni and nc-Ni samples at 3 K and 300 K: coercive field ($H_c$) and magnetic field ($H_p$) at the peak position of the maximum/minimum of the MR(H) curves.*

| sample | $H_c$ (Oe) | | $H_p$ (Oe) | |
|---|---|---|---|---|
| | 3 K | 300 K | 3 K | 300 K |
| µc-Ni (#B5) | ca. 10 | 13 | 85 | 75 |
| nc-Ni (#B2) | ca. 45 | 37 | 140 | 160 |

The coercive field $H_c$ of the µc-Ni sample matches well the usual coercive fields known for well-annealed bulk Ni [3]. The coercive field of the nc-Ni sample compares well with the results of Aus et al. [47] who reported at room-temperature coercive fields between 25 and 40 Oe for electrodeposited nc-Ni with grain sizes ranging from 10 to 1000 nm, both in the as-deposited and annealed states. Kisker et al. [49] reported room-temperature coercive fields of the order of 100 Oe for nc-Ni with 10 nm grain size which was produced by a careful compaction of gas-phase-condensed Ni nanoparticles. These latter values are larger than our ones for nc-Ni by a factor of two, but this may be due to the quite different preparation methods and, thus, the resulting microstructures since the coercive field strongly depends on the microstructure.

In any case, by looking at the coercive field data in Table II, we have the general relation $H_c$(µc-Ni) < $H_c$(nc-Ni) at both temperatures. The larger coercive field of nc-Ni may partly arise due to the residual stress in the as-deposited nc-Ni foil as typical for layers produced via atom-by-atom deposition processes. Such residual stress can then give rise to a stress-induced magnetic anisotropy since the magnetostriction of Ni is not zero and this results in higher coercivity. On the other hand, the small $H_c$ value for μc-Ni is understandable due to the fact that this sample is a well-annealed, defect-free foil which is in a stress-free state and, therefore, a stress-induced magnetic anisotropy does not arise.

It should also be noted that at both temperatures the relative remanence was at least 50 % for the nc-Ni sample whereas the well-annealed μc-Ni sample exhibited a definitely much lower relative remanence. The well-defined large relative remanence of the nc-Ni sample is another indication that it may have a substantial in-plane magnetic anisotropy due to, e.g., the residual stresses. In contrast, the lack or low value of the relative remanence for the μc-Ni sample speaks for a low in-plane magnetic anisotropy. These features well support the obtained difference in the coercive fields of the two microstructural states of Ni.

In the MR($H$) curves, the hysteresis behavior is characterized by the field values ($H_p$) of the magnetoresistance minimum (LMR) and maximum (TMR) peaks. The $H_p$ data obtained are also collected in Table II which reveal a relation $H_p$(μc-Ni) < $H_p$(nc-Ni). This matches well the relation $H_c$(μc-Ni) < $H_c$(nc-Ni) obtained above for the coercive fields when comparing the microcrystalline and nanocrystalline states. On the other hand, for both microstructural states, we have the relation $H_c$ < $H_p$. As we can infer from Table II for the nc-Ni sample, the coercive force ($H_c$) and the peak position of the MR($H$) curve ($H_p$) are different by a factor of about 3 to 4 and the difference is apparently even larger for μc-Ni.

It is not straightforward to rationalize the relation $H_c$ < $H_p$ and in the following, we will attempt only to illuminate at least why the two hysteresis parameters can be different. The two quantities ($H_c$ and $H_p$) reflect critical magnetic field points of the remagnetization process where the distribution of the domain magnetization orientations exhibits an extremum. $H_c$ and $H_p$ are usually close to each other although they are usually not equal. This is because the conditions for zero magnetization ($H = H_c$) during the magnetization reversal are not identical with the conditions for the occurrence of maximum/minimum resistivity peaks during cycling the external magnetic field.

At $H = H_c$, the field-direction-projected domain magnetization components aligned parallel and antiparallel to the magnetic field orientation sum up to zero. At the field position of the LMR minimum, $H_p$(LMR), the absolute values of the field-direction-projected magnetization components, irrespective of whether they are aligned parallel or antiparallel to the magnetic field orientation, sum up to a minimum value. Since the same domain magnetizations can be split into components parallel (LMR) and perpendicular (TMR) to the magnetic field direction (not to the magnetic field orientation), the TMR maximum, $H_p$(TMR), should appear at the same magnetic field as the LMR minimum. This was well demonstrated in Fig. 13b with the MR($H$) data of the nc-Ni sample at $T = 300$ K where the accuracy of the measured resistivity data and the shape of the MR(H) curves enabled us to observe the validity of the relation $H_p$(LMR) = $H_p$(TMR).

As to the temperature dependence, the coercive field is usually known to increase with decreasing temperature what follows from the inherent temperature dependence of the magnetocrystalline anisotropy. Although the limited accuracy of the coercive field determination by our SQUID equipment for such low $H_c$ values does not allow us to clearly see the usual temperature dependence of the coercivity, but the results at the two temperatures are reasonable. The situation is very similar for the other coercivity parameter $H_p$ which does not change significantly with temperature for either of the two samples.

*4.4 Comparison of the magnetoresistance of µc-Ni and nc-Ni at T = 300 K*

A comparison of Figs. 10a and 13a reveals that the magnetic-field-induced resistivity changes at $T = 300$ K are qualitatively the same for the µc-Ni and nc-Ni samples, just the values of the derived magnetoresistance parameters are slightly different. As demonstrated in Figs. 12 and 15, the variation of the resistivity with magnetic induction could be properly described for the measured data in the magnetically saturated (monodomain) state by a second-order polynomial empirical fitting function: $\rho(B) = \rho(B=0) + \alpha \cdot B + \beta \cdot B^2$. For convenience, the values of the parameters $\alpha$ and $\beta$ as well as the $\rho(B=0)$ values are collected in Table III where the very close values of the fits to the $\rho(H)$ vs. $H$ plots are also included (in brackets).

*Table III  Parameters characterizing the field dependence of the magnetic-field-induced resistivity change $\Delta\rho(B) = \rho(B) - \rho(B=0)$ of the µc-Ni and nc-Ni samples at 300 K as derived in Figs. 12 and 15. The parameter values were obtained as explained in the text from fits of the experimental resistivity data in the magnetically saturated region to the empirical function $\rho(B) = \rho(B=0) + \alpha \cdot B + \beta \cdot B^2$. The values in brackets provide the fit parameters for the $\rho(H)$ vs. H plots.*

| $T = 300$ K fitted parameters | | $\rho(B=0)$ (µΩ·cm) | $\alpha$ ($10^{-3}$·µΩ·cm/kG) | $\beta$ ($10^{-6}$·µΩ·cm/(kG)$^2$) |
|---|---|---|---|---|
| µc-Ni (#B5) | LMR | 7.490 (7.481) | -1.464 (-1.436) | +2.272 (+2.272) |
| | TMR | 7.318 (7.309) | -1.578 (-1.544) | +2.791 (+2.791) |
| nc-Ni (#B2) | LMR | 8.893 (8.885) | -1.340 (-1.326) | +1.165 (+1.162) |
| | TMR | 8.725 (8.715) | -1.535 (-1.510) | +2.028 (+2.028) |

In order to better visualize the very similar behavior of the field dependence of the resistivity for the two microstructural states of Ni, the magnetic-field-induced resistivity change $\Delta\rho(B) = \rho(B) - \rho(B=0)$ is displayed in Fig. 16 (the fitting functions are only shown) as a function of the magnetic induction $B$ for both samples. It can be established that for each microstructural state, the TMR component is more negative than the LMR component whereas for a given component (LMR or TMR), the data practically agree for the µc-Ni and nc-Ni samples, the agreement being especially good for the TMR component. All these mean that for a given MR component, the magnetic-field-induced resistivity change is the same for both the microcrystalline (bulk) state and the nanocrystalline state (at least for the presently investigated 100 nm grain size). Furthermore, for both microstructural states, the resistivity change is definitely larger for the TMR configuration than for the LMR configuration.

For an explanation of the recent two features (close quantitative agreement of $\Delta\rho$ for µc-Ni and nc-Ni as well as the relation $|\Delta\rho(TMR)| > |\Delta\rho(LMR)|$ for a given microstructural state), we can proceed along the following considerations. Due to the strong electron-phonon interaction at $T = 300$ K, the mean free path in Ni at this temperature is very short ($\lambda$(Ni,300K) = 5.5 nm [41]). Therefore, the majority of the electron scattering events, even in the nc-Ni sample with an average grain size of about 100 nm, occur within the crystallites and only a small fraction of scattering events takes place at the grain boundaries. As a consequence, for a given magnetic field configuration with respect to the current flow direction (LMR or TMR), the influence of magnetic field on the spin-fluctuation-suppression process is not expected to be noticeably different for the two microstructural states. It may also be that this suppression process does not differ for scattering events within the crystallites and in the grain boundaries. As to the second feature, the stronger dependence of resistivity on induction for the TMR configuration may be simply a consequence of the stronger suppression influence of the induction for the transverse spin fluctuations than for the longitudinal ones. A proper theoretical description should take into account also this possibility in order to explain the observation that $|\Delta\rho(TMR)| > |\Delta\rho(LMR)|$ and this will be further discussed in Section 5.

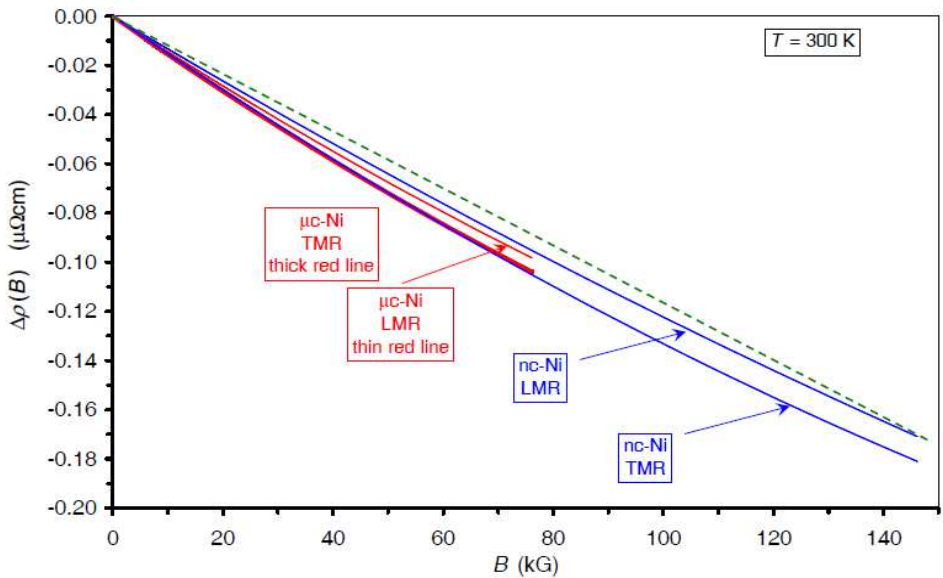

*Figure 16* *Field-induced resistivity change $\Delta\rho(B)= \rho(B) - \rho(B=0)$ vs. B at T = 300 K for $\mu$c-Ni and nc-Ni with magnetic field orientations as indicated (LMR, TMR). For the sake of clarity, the fitting functions with parameters listed in Table III are only displayed. The dashed straight line connecting the B = 0 and B = 146.1 kG data points for the LMR data of the nc-Ni sample was only added to better illustrate the general deviation (upward curvature) of the $\Delta\rho(B)$ data from the linear behavior.*

In contrast to the data for $T = 3$ K, we do not display the room-temperature results in the form of the relative change $\Delta\rho(B)/\rho(B=0)$ with magnetic induction since, due to the similar field evolution of the resistivity change for the two samples, the same picture is obtained as shown in Fig. 16 except that the relative change is somewhat smaller for nc-Ni due to its higher zero-induction resistivity as a consequence of the presence of a large amount of grain boundaries.

It is also noted that, as discussed in the previous two subsections, the behavior of the field-induced resistivity change at $T = 300$ K would be practically the same if we display the data as a function of the applied magnetic field since $\rho(H=0)$ differs from $\rho(B=0)$ by at most 0.1 % due to the negligible OMR contribution at such a high temperature [34].

Adjoining this last issue, a further remark is added here about the evaluation of the AMR ratio as usually done by displaying the relative resistance change $\Delta R/R_o = [R(H) - R_o]/R_o$, which is equivalent to $\Delta\rho/\rho_o = [\rho(H) - \rho_o]/\rho_o$, as a function of the magnetic field $H$. If there is no or at most a negligible hysteresis of the MR($H$) curve (see, e.g., Fig. 2b for both LMR and TMR and Fig. 10b for the TMR component), then, evidently, the maximum or minimum value of the resistivity can be taken as $\rho_o$. However, Figs. 5b, 10b and 13b reveal a significant hysteresis of the MR($H$) curve in which cases the zero-field resistivity $\rho(H=0)$ is definitely different from the $\rho(H_p)$ value. As a consequence, when displaying the relative resistivity change as a function of the magnetic field, i.e., $\Delta\rho/\rho_o$ vs. $H$, it is important whether the zero-field (or "remanent") resistivity $\rho(H=0)$ or the peak resistivity $\rho(H_p)$ is taken as reference value $\rho_o$ since it will influence the value of AMR ratio. For example, according to Fig. 10b, if we take $\rho(H=0)$ as reference value for the LMR component [for the TMR component $\rho(H=0) \approx \rho(H_p)$], we get about 1.55 % for the AMR ratio whereas if we take $\rho_o = \rho(H_p)$, then we get about 2.35 %. Intuitively, the common practice has always been to take $\rho_o = \rho(H_p)$ because from magnetoresistance point of view the really "demagnetized" state is at the smallest (LMR) or largest (TMR) resistivity values around zero.

In Section 4, we were able to deduce the room-temperature AMR ratio with the help of the exact Kohler analysis on the basis of the accurately measured resistivities as a function of the magnetic induction $B$ with values AMR($\mu$c-Ni) = 2.35 % and AMR(nc-Ni) = 1.91 %. In this analysis, we were relying on the high-field resistivity data only, i.e., the uncertainty due to the low-field hysteresis is eliminated. By analyzing the same resistivity data in the form of $\Delta\rho/\rho_o$ vs. $H$ with the choice $\rho_o = \rho(H_p)$, we obtained AMR($\mu$c-Ni) = 2.36 % and AMR(nc-Ni) = 1.98 %. These excellent agreements justify that the peak resistivity is the proper choice as reference value when determining the AMR ratio from a $\Delta\rho/\rho_o$ vs. $H$ plot.

# 5. Analysis of the field dependence of the magnetoresistance of μc-Ni and nc-Ni in the magnetically saturated state at $T = 300$ K

*5.1 Field dependence of the resistivity in the saturation region in Ni at $T = 300$ K: comparison with previous experimental results*

The data in the previous section revealed that in the magnetically saturated (monodomain) state, the field dependence of the resistivity at $T = 300$ K is very similar for the two microstructural states of Ni (μc-Ni and nc-Ni). As already discussed earlier, the resistivity decrease at high temperatures is due to the suppression of the thermally-induced magnetic disorder [4,22,33,34] by the external magnetic field because this magnetic disorder gives a contribution to the zero-field resistivity in addition to the contributions due to the static and dynamic topological disorder (lattice defects including impurities and lattice vibrations, respectively). An increasing external magnetic field reduces the magnetic disorder (paraprocess) and, therefore, the resistivity contribution due to conduction electron scattering on the non-aligned magnetic moments is diminished. In not too high magnetic fields, this field-induced resistivity change is nearly linear [24,39,40]. In large magnetic fields, in agreement with reported literature results [22,33,34], we have also observed a slight upward curvature of the MR($H$) curves (cf. Figs. 12 and 15). For our two Ni samples (both μc-Ni and nc-Ni), the field dependence observed at room temperature could be accurately described phenomenologically by a second-order polynomial with parameters collected in Table III above.

In the following, we will compare the field dependence of the induced resistivity change observed in our samples in the saturated region with relevant available earlier results [22,33,34]. Madduri and Kaul [22] investigated the field and temperature dependence of the resistivity of electrodeposited nc-Ni samples with XRD crystallite sizes from 10 nm to 40 nm. The field-induced resistivity change was measured in the TMR configuration up to $H = 90$ kOe from 2 K to 300 K and the temperature dependence of the zero-field resistivity was also reported for these samples. In zero magnetic field, the room-temperature resistivity and the residual resistivity as well as the RRR ratio of their nc-Ni sample with 40 nm crystallite size were well comparable with the corresponding parameters of our nc-Ni sample with 75 nm crystallite size.

Raquet et al. [33,34] measured the resistivity of an MBE-grown patterned Ni film (20 nm thickness and deposited on a MgO substrate) from 77 K to 443 K up to $B$ = 300 kG in the LMR configuration with the magnetic field in the film plane. No details about the structure of their Ni film were given and the absolute value of the resistivity was also not provided, only the field-induced resistivity change $\Delta\rho(B)$ was reported. The only note about sample characterization is that "a residual resistance ratio around 27 for the thicker films attests to their high structural quality" [34]. In the same report, these authors investigated 7-nm-thick Co films, 20-nm-thick Ni films and 80-nm-thick Fe films and it is not clear for which film the specified RRR = 27 refers to. However, even if we assume that their 20-nm-thick Ni film exhibited this RRR value, this does not testify for their Ni sample a high film quality in terms of impurities and lattice defects. The specified RRR = 27 value corresponds to a residual resistivity of about 0.25 $\mu\Omega\cdot$cm by assuming the bulk resistivity value for their Ni film at room temperature. This residual resistivity is of the same order of magnitude as the value for our nc-Ni sample or in the work of Madduri and Kaul [22]. Therefore, the Ni film studied by Raquet et al. [34] can also be considered as a nanophase Ni sample (at such small thicknesses, thin films are typically nanocrystalline).

In Fig. 17, we show a comparison of the high-field resistivity data for our nc-Ni sample at $T$ = 300 K for both the LMR and TMR components together with the LMR result of Raquet et al. [34] for the 20-nm-thick Ni film at $T$ = 284 K (the closest temperature to our measurement temperature) and with the TMR results of Madduri and Kaul [22] for two of their nc-Ni samples at $T$ = 300 K. The $\Delta\rho(B)$ data from the three independent reports on nanophase Ni samples match each other relatively well and, with reference to Fig. 16, also the corresponding data on our $\mu$c-Ni sample (bulk Ni) for which the data were omitted from Fig. 17 for the sake of clarity. Although the slopes of the $\Delta\rho(B)$ curves are somewhat different quantitatively for the samples compared in Fig. 17, the difference amounting to about a factor of 2 up or down with respect to our data, the deviation of the $\Delta\rho(B)$ data from a linear behavior (upward curvature) can be clearly established for each sample investigated up to now.

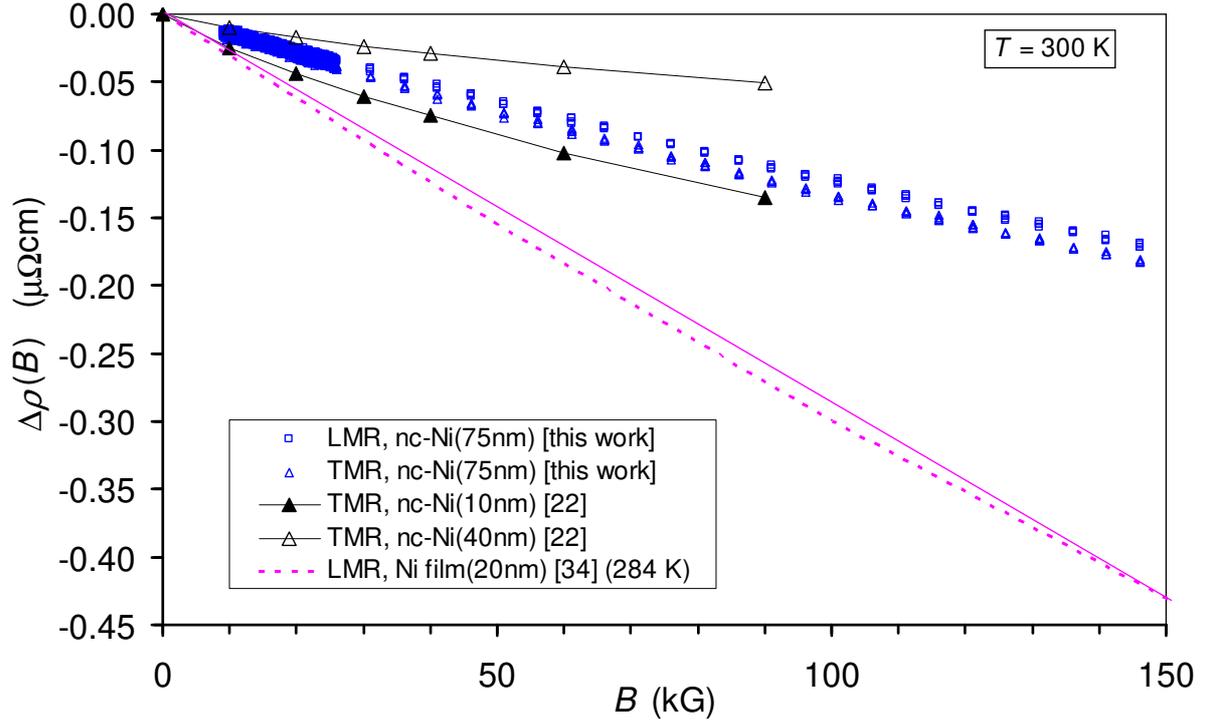

***Figure 17*** *Field-induced resistivity change Δρ(B)= ρ(B) – ρ(B=0) vs. B at T = 300 K for our nc-Ni sample with magnetic field orientations as indicated (LMR, TMR) in comparison with available results on other nanophase Ni samples [22,34]. For the nc-Ni samples, the XRD crystallite sizes are given in the brackets whereas the film thickness is given for the Ni film. In Ref. 34, the data displayed here by the dashed line extend up to 300 kG with the same trend. The straight solid line connecting the B = 0 and B = 150 kG data points of Ref. 34 was only added to better illustrate the deviation (upward curvature) of the Δρ(B) data from the linear behavior.*

It should be noted furthermore that whereas the data obtained on massive electrodeposited nc-Ni samples in the present work and in Ref. 22 are fairly close to each other, the slope for the 20-nm-thick Ni thin film sample [34] is definitely much larger. The reason for this discrepancy may arise from the fact that in a Ni thin film an additional term to the resistivity may stem from surface scattering effects [51-53]. Due to the short mean free path at around 300 K, in zero magnetic field this resistivity contribution may be negligible since the film thickness (20 nm) is about four times larger than the mean free path (5.5 nm). However, in a magnetic field the electrons do not only randomly move towards lower electric potentials, but during their free travel they are also often deflected from this direction due to the Lorentz force, making scattering events at the nearby film surface also possible which then causes an additional resistivity contribution in a thin film with respect to a massive sample, even having both the same microstructure (e.g. the same grain size). Therefore, we can see that helical motion of electrons due to the Lorentz force further magnifies this zero-field surface (i.e., finite-size) scattering effect in high magnetic fields.

### *5.2 Analysis of the field dependence of the resistivity in the monodomain region in Ni metal around room temperature in view of previously proposed theoretical models*

We turn now attention to the results of theoretical considerations put forward in Refs. 22, 34 and 54 on the evolution of the resistivity with magnetic field in the magnetically saturated (monodomain) state of ferromagnetic metals around room temperature. First, we discuss the results of Raquet et al. [34] who addressed the determination of the collective excitations and their contributions to the intrinsic resistivity via spin-flip scattering for 3d ferromagnets. They demonstrated that this almost linear, non-saturating high-field negative magnetoresistance has a magnetic origin and it was assigned to the suppression of electron-magnon scattering and the damping of spin waves with increasing magnetic field. These authors carried out a theoretical calculation of the magnetic resistivity originating from spin-flip intraband *s-s* and *d-d* as well as interband *s-d* transitions via electron-magnon diffusion including both the high-field effect on the magnon spectrum and the magnon mass renormalization. By restricting the analysis to the temperature range $T_C/5 < T < T_C/2$ and to magnetic fields $B < 1000$ kG, they arrived at the following approximate analytical expression for the field dependence of the induced resistivity change:

$$\Delta \rho(T,B) \approx \rho(T,B) - \rho(T,B=0) = A_1 \, [B \, T/D(T)^2 \, \ln[\mu_B B/(k_B T)]] \qquad (1)$$

where $A_1$ is a proportionality constant, $D(T)$ the spin wave stiffness constant at temperature $T$, $\mu_B = 0.927 \cdot 10^{-20}$ erg/G the Bohr magneton and $k_B = 1.38 \cdot 10^{-16}$ erg/K the Boltzmann constant. Raquet et al. [34] claimed that their experimental data for a Ni(20nm) thin film can be well described by this expression for temperatures not exceeding 320 K and for magnetic fields below 200 kG whereas for higher temperatures and larger magnetic fields, the theoretical curves exhibited a significantly larger upward curvature than the experimental data. Since we wish to analyze the field dependence of our Ni samples at $T = 300$ K which is just below $T_C(\text{Ni})/2 \approx 315$ K and our highest magnetic field does not exceed 150 kG, eq. (1) can be tested against the present experimental data for µc-Ni and nc-Ni metal samples.

To proceed along this line, for convenience, eq. (1) is rewritten in the following form:

$$\Delta\rho(T,B) \approx A_1 (k_B/\mu_B) [T^2/D(T)^2] [\mu_B B/(k_B T)] \ln[\mu_B B/(k_B T)]. \tag{2}$$

By introducing the notations $A_2 = A_1 (k_B/\mu_B) [T^2/D(T)^2]$ and $x = \mu_B B/(k_B T)$, we finally get a simple expression for the field and temperature dependence of the field-induced resistivity change:

$$\Delta\rho(T,B) \approx A_2 \, x \ln(x). \tag{3}$$

As we could see in Fig. 16 that at $T = 300$ K, the $\Delta\rho(B)$ vs. $B$ data were qualitatively the same and also quantitatively very close to each other for both the LMR and TMR components of our µc-Ni and nc-Ni samples, we will analyze only the data for the TMR component of the nc-Ni sample investigated in a wider magnetic field range. By taking into account that in our case $T = 300$ K and in the magnetic saturation range we have for the magnetic field the relation 3 kOe $\leq H \leq$ 140 kOe (i.e., 9.1 kG $\leq B \leq$ 146.1 kG), we get that the range of possible values of $x$ is $0.001367 \leq x \leq 0.032749$.

Our experimental $\Delta\rho(B)$ vs. $B$ data (□) and their fit (thin line) to the empirical function $\rho(B) = \rho(B=0) + \alpha \cdot B + \beta \cdot B^2$ with parameters as given in Table III are displayed in Fig. 18. The symbols ◇ represent eq. (3) with parameter $A_2 = 1.55$ which value gave the best agreement with the experimental data. It can be clearly seen that there is a systematic deviation in that eq. (3) overestimates and underestimates the low-field and high-field experimental data, respectively (if we try to match the experimental data at low fields with a somewhat smaller $A_2$ value, the deviation will be much larger at high fields). The fit result was the same for the LMR component of our nc-Ni sample and also for both the LMR and TMR components of the µc-Ni sample. It should be noted that even the experimental data of Raquet et al. [34] for $T = 284$ K (displayed in Fig. 17) show the same deviation from the theoretical values given by eq. (3) as observed for our samples.

The other recent attempt to describe theoretically the effect of the suppression of spin fluctuations on the resistivity of ferromagnetic metals has been made by Kaul [54]. In order to go beyond the electron-gas approximation used by Raquet et al. [34] which is definitely not a valid approach for $3d$ transition metals, Kaul [54] carried out, in the framework of the two-band (s and d bands) model, a self-consistent calculation of the spin-fluctuation contributions to the resistivity in the absence and presence of a magnetic field for a weak itinerant-electron ferromagnet (WIF) and the model was successfully applied to explain the magnetotransport properties of some WIF materials [55].

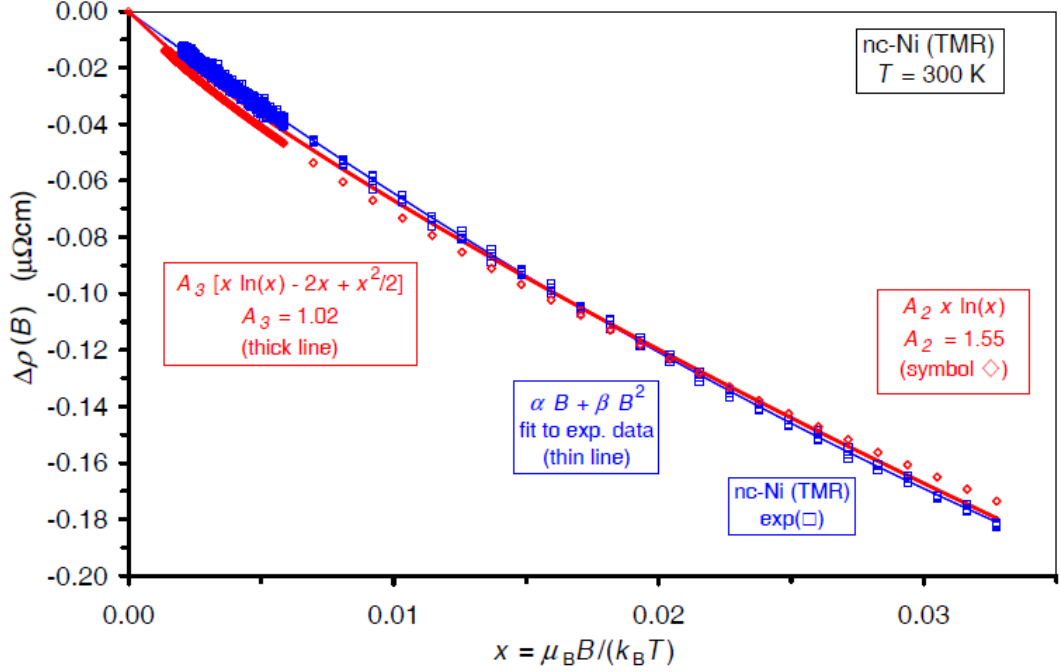

**Figure 18** *Comparison of the experimental data (□) of the field-induced resistivity change $\Delta\rho(B)$ at $T = 300$ K for the TMR component of our nc-Ni sample with some theoretical formulae suggested for the field dependence of $\Delta\rho(B)$. The symbols ◇ denote the theoretical $\Delta\rho(B)$ values according to eq. (3) with a value of $A_2$ to be as close as possible to the experimental data. The thick line denotes theoretical $\Delta\rho(B)$ values according to eq. (5) with a value of $A_3$ to be as close as possible to the experimental data. The thin line denotes a second-order polynomial fit to our experimental data.*

The overall conclusion in Ref. 54 was that the analytical expressions derived in various temperature ranges for the spin-fluctuation contribution to the field-induced resistivity change basically have a simple general form

$$\Delta\rho(B) \propto [a\,B + b\,B^2] \qquad (4)$$

with $a < 0$ and $b > 0$. This corresponds to our empirical fit function $\Delta\rho(B) = \rho(B) - \rho(B=0) = \alpha\,B + \beta\cdot B^2$ which was found to describe with high accuracy the field dependence of the resistivity at $T = 300$ K for our μc-Ni and nc-Ni samples (cf. Figs. 12 and 15).

Madduri and Kaul [22] have applied later the model elaborated for WIF metals [54] for explaining the field and temperature dependence of the magnetoresistance of their nc-Ni samples in spite of the fact that Ni is a strong itinerant-electron ferromagnet (SIF). In particular, they picked up a specific result from Ref. 54 according to which for $T \ll T_C$ the field-dependence of the resistivity due to the suppression of the thermally excited spin fluctuations (spin waves) can be described as

$$\Delta\rho(T,B) \approx A_3 \, [x \ln(x) - 2\,x + x^2/2] \qquad (5)$$

where $A_3 = \eta \, \rho(T,B=0)$ with $\eta = 0.304$.

Madduri and Kaul [22] claimed that the experimental data for their nc-Ni samples could be closely reproduced by using the leading term $x \ln(x)$ only in eq. (5) for temperatures 2 K < $T$ < 300 K in magnetic fields up to 90 kOe. Instead of the fixed prefactor $\eta = 0.304$ in eq. (5), the prefactor was also allowed to be a free variable to achieve good quantitative fits with $\eta$ values ranging from 0.11 to 0.43 for the nc-Ni samples with crystallite sizes ranging from 10 nm to 40 nm.

We have tested eq. (5) to our data and the best approach is shown by the thick line with $A_3 = 1.02$ in Fig. 18. Since according to Table III, $\rho(T,B=0) = 8.725$ μΩcm for the TMR component of our nc-Ni sample, by using the fit value $A_3 = 1.02$, we get $\eta = 0.12$ for our nc-Ni sample which corresponds to the $\eta = 0.11$ value of Madduri and Kaul [22] for their nc-Ni sample with a crystallite size of 10 nm.

By considering the data according to eq. (3) in Fig. 18, we should establish that, in contrast to the results of Madduri and Kaul [22], the leading term $x \ln(x)$ alone does not give a satisfactory agreement with experiment for our nc-Ni sample (and also not for our μc-Ni sample) since there is a systematic deviation from the experimental data. Even the full form of eq. (5), although it leads to a better agreement with experiment (see Fig. 18), shows a similar systematic deviation as eq. (3), just to a lesser extent. Clearly, our empirical function corresponding to eq. (4) gives a significantly more accurate description of the experimental data than the available theoretical approaches [22,34,54] represented either by eqs. (3) or (5).

This also implies that there is space for a refinement of the theoretical description of the suppression of the spin-fluctuation contribution to the field-induced resistivity change in the saturated states of ferromagnetic metals at high temperatures. Specifically, an explanation of the stronger field dependence of the TMR component than that of the LMR component is still lacking.

# 6. Correlation between the zero-field resistivity and the resistivity anisotropy splitting parameter $\Delta\rho_{AMR}$ in Ni metal

In a recent paper [24] devoted to the room-temperature study of a series of electrodeposited nc-Ni samples with various grain sizes in magnetic fields up to $H = 9$ kOe, we have found a linear correlation between the resistivity anisotropy splitting parameter $\Delta\rho_{AMR}$ and the resistivity $\rho_o$ measured in the absence of a magnetic field with the trendline describing the correlation going through the origin. Actually, this correlation is not so surprising since the AMR ratio is just defined as the ratio of these two transport parameters. However, the interesting point in that study [24] was that the anisotropic magnetoresistance ratio AMR = $\Delta\rho_{AMR}/\rho_o$ remained constant if we varied the zero-field resistivity by changing the grain size in nc-Ni, i.e., incorporating various density of lattice defects into the Ni lattice whereby practically unaffecting the bulk properties of Ni in terms of the electronic density of states around the Fermi level. [We refer here to the theoretical work of Szpunar et al. [50] who demonstrated that the features of the band structure and the magnitude of the magnetic moment of Ni atoms in the grain boundaries are not significantly different from Ni atoms in the crystallite interior.] According to the above finding in the present paper, there is no anisotropy splitting at $T = 0$ K where the resistivity vanishes in pure bulk Ni whereas when the resistivity increases, e.g., due to lattice vibrations at higher temperatures, there will be a resistivity anisotropy splitting $\Delta\rho_{AMR}$ in proportion to the actual $\rho_o$ value within such a series as long as the Fermi-level density of states of electrons remains unchanged.

Having now the results of the high-field magnetoresistance measurements at $T = 3$ K and $T = 300$ K of one nc-Ni and one µc-Ni sample of the previous study [24], we can extend the $\Delta\rho_{AMR}$ vs. $\rho_o$ correlation (Fig. 4 of Ref. 24) by including the present new data as shown in Fig. 19. We can see that the room-temperature results of the present work on the µc-Ni and nc-Ni samples agree well with data of the previous study [24] on the same samples. The solid line in Fig. 19 indicates the trendline obtained from the magnetotransport data of the series of nc-Ni samples with various grain sizes in Ref. 24. As demonstrated in Ref. 24, several bulk Ni and nc-Ni samples both from our measurements and from literature reports follow (with some scatter) the trendline over the data of the nc-Ni series which yielded AMR = 2.39 %. Both the old and new results on the µc-Ni sample (#B5) comply with the trendline.

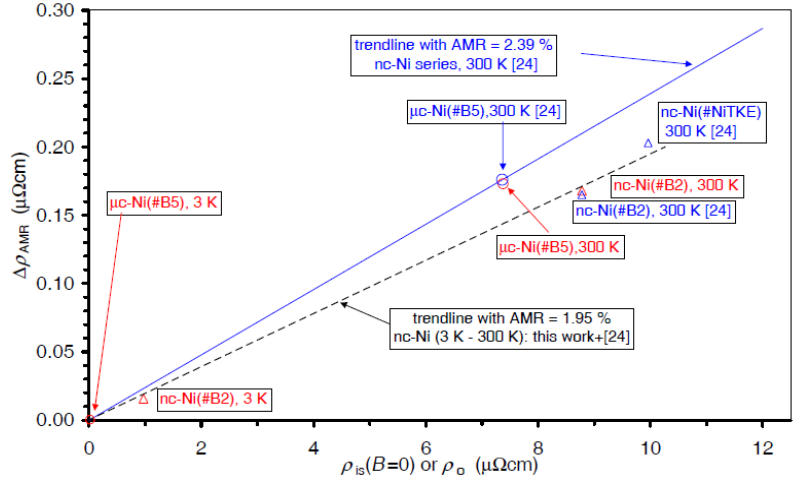

***Figure 19*** *Correlation of the resistivity anisotropy splitting $\Delta\rho_{AMR}$ with the zero-induction isotropic resistivity $\rho_{is}(B=0)$ or resistivity $\rho_o$ measured in the absence of a magnetic field for various Ni samples at temperatures between T = 3 K and 300 K. Color codes of data: red (current results); blue (results from Ref. 24). Symbols ○ stand for bulk/microcrystalline (µc) Ni; symbols △ stand for nc-Ni samples. Solid (blue) line: trendline from Ref. 24 for data on a series of electrodeposited nc-Ni samples with various grain sizes; dashed (black) line: trendline for all the magnetotransport measurements (this work and Ref. 24) on two electrodeposited nc-Ni samples produced in a much earlier work [56]. Attached to each data point is a text box specifying the sample code and measurement temperature as well as data source if not from this work. According to the results in previous sections, the difference between the zero-field and zero-induction resistivity is negligible at room temperature, but $\rho_{is}(B=0)$ should be used at low temperatures.*

As to the results on the nc-Ni sample (#B2) from this work and Ref. 24 as well as on the nc-Ni sample #NiTKE from Ref. 24, these data exhibit a systematic downward deviation form the trendline corresponding to AMR = 2.39 %. These two particular samples were prepared more than two decades ago in a previous study [56] under different electrodeposition conditions than the samples of the nc-Ni series in Ref. 24. Nevertheless, the room-temperature and low-temperature data of these two samples also define a common trendline (dashed line in Fig. 19) corresponding to AMR = 1.95 % so their qualitative behavior is the same as for the rest of all Ni samples included in Fig. 19. An explanation for the lower apparent AMR ratio of these two nc-Ni samples may be the eventual incorporation of a small concentration of non-metallic elements (C, O and S) which may have remained unrevealed by our SEM EDX analysis as discussed in Refs. 23 and 24; on the basis of the specific electrodeposition conditions (especially concerning the bath composition) in Ref. 56, this cannot be completely excluded. If such impurities are present in Ni, they can increase the resistivity and, at the same time, may also cause a slight change of the electronic density of states resulting in a decrease of the resistivity anisotropy splitting. Actually, both effects tend to cause just the kind of deviation as we see in Fig. 19 with respect to the trendline defined by the certainly much purer new nc-Ni series samples of Ref. 24 and also by results on several bulk or μc-Ni samples.

In summary, we can conclude that for a given series of Ni samples with a common level of impurity content, we observe a linear correlation between the resistivity anisotropy splitting parameter $\Delta\rho_{AMR}$ and the zero-field resistivity $\rho(H=0)$ when varying either the grain size or the temperature. The correlation then defines a given AMR ratio that may depend on the impurity content.

**7. Summary**

In this work, the field evolution of the resistivity was studied at $T = 3$ K and 300 K on a μc-Ni sample with large grain sizes (corresponding to bulk Ni) in magnetic fields up to $H = 70$ kOe and on a nc-Ni sample with an average grain size of about 100 nm up to $H = 140$ kOe. In order to relate the variation of the resistivity with magnetic field to the magnetization process, the magnetization isotherms were also measured at both temperatures. The main features of the magnetization process (hysteresis and approach to saturation) were found to show up in most cases also in the MR($H$) curves.

The basic purpose of this study was to establish if the magnetoresistance behavior of Ni metal changes with respect to the bulk state by introducing a large density of grain boundaries as two-dimensional scattering centers. The presence of grain boundaries yields an increase of the resistivity [18,23,26] at any temperature as a result of which the residual resistivity becomes substantial for the nc-Ni state (amounting to about 11 % of the room temperature value for the present nc-Ni sample) as opposed to the vanishingly small residual resistivity of the microcrystalline state. This difference results, at the same time, in a great difference of the electron mean free paths between the two microstructural states at low temperatures and this showed up in the different MR behaviors of the two samples. This is mainly because at low temperatures in pure bulk metals, as a consequence of the long mean free path, the Lorentz-force-induced contribution (OMR) to the resistivity in a magnetic field is very large whereas for a nanocrystalline metal the shorter mean free path effectively reduces the OMR contribution, although cannot completely eliminate it.

At low temperature ($T$ = 3 K), the MR($H$) behavior of microcrystalline (bulk) Ni corresponded to previous reports in that in the magnetically saturated state, the resistivity increased with magnetic field. The present high-precision resistivity data revealed that both the LMR and TMR components exhibit a tendency for saturation towards high magnetic fields. For the μc-Ni sample, both the TMR and LMR components increased continuously immediately starting from $H$ = 0 until the largest field applied and the relation TMR($H$) > LMR($H$) persisted in the whole range of magnetic fields investigated. An interesting point is that at the magnetic field where the monodomain state (technical saturation) is achieved according to the measured $M(H)$ curves, one can hardly see any specific feature in the MR($H$) curves of the μc-Ni sample,

At $T$ = 300 K, the MR($H$) curves of the μc-Ni sample were similar to those known for bulk Ni. After magnetic saturation, the resistivity decreased nearly linearly (with a slight upward curvature) with magnetic field which variation is due to the suppression of thermally-induced magnetic disorder with increasing magnetic field.

By contrast, for nc-Ni at $T$ = 3 K, in magnetic fields below saturation, both MR components exhibited a similar field dependence as observed for bulk Ni at 300 K (LMR > 0, TMR < 0). However, above magnetic saturation, both components started to increase without any sign of saturation up to the highest magnetic fields applied at $T$ = 3 K. The rate of increase with field was stronger for the TMR component, so that above around 90 kOe, TMR became larger than LMR. On the other hand, the rate of the relative resistivity change for the nc-Ni sample at $T$ = 3 K was by about an order magnitude smaller than for the μc-Ni sample that could be explained by the different electron mean free paths at T = 3 K for the two microcrystalline states. At $T$ = 300 K, the MR($H$) curves of nc-Ni were qualitatively the same as for μc-Ni and the rate of decrease with magnetic field was practically the same for both samples and somewhat smaller for the LMR component than for the TMR component.

The MR($H$) data in the magnetically saturated state were analyzed at both temperatures with the help of Kohler plots from which the zero-induction resistivities could be deduced for both the LMR and TMR components (at room temperature, these resistivities agreed closely with the zero-field values). Based on these values, the experimental data on the dependence of the resistivity on magnetic induction (and magnetic field) could be very accurately fitted for both samples with an empirical fourth-order (μc-Ni at $T$ = 3 K), third-order (nc-Ni at $T$ = 3 K) and second-order (both μc-Ni and nc-Ni at $T$ = 300 K) polynomial.

The room-temperature data on our samples were compared to available experimental results in previous reports and good qualitative agreement for the field dependence of the resistivity in the saturation (monodomain) range was found, although the strength of the field dependence showed a scatter by a factor of 2 between our samples and previous reports. We have also tested the previously suggested theoretical field dependence of the suppression of the spin-fluctuation contribution to the resistivity and it could be established that the empirical second-order polynomial function gives a much better description of the experimental data than the previously suggested model formulae. This clearly invokes for a refinement of the theoretical description of the field dependence of the spin-fluctuation resistivity contribution, especially concerning the observed stronger field dependence for the TMR component in comparison with the LMR component.

The Kohler-plot analysis results have also enabled us to derive the resistivity anisotropy splitting ($\Delta\rho_{AMR}$) and the anisotropic magnetoresistance (AMR) ratio. For the µc-Ni sample, the AMR ratio at $T = 3$ K could be determined with a large uncertainty only as $(2.25 \pm 1.10)$ % and this value is quite close to the room-temperature value of $(2.35 \pm 0.03)$ %. The magnetotransport parameters deduced for the µc-Ni sample here fit well into the trend obtained recently [24] for bulk and nc-Ni metal samples with a correlation between the $\Delta\rho_{AMR}$ data and the zero-field resistivities.

For the nc-Ni sample, the AMR ratio at $T = 3$ K could be determined with much better accuracy as $(1.62 \pm 0.04)$ % which is by about 15 % smaller than the room-temperature value of $(1.91 \pm 0.03)$ %. Nevertheless, the data for this nc-Ni sample, together with the room-temperature result for another nc-Ni sample from the same old batch [24], exhibit a similar trend on a $\Delta\rho_{AMR}$ vs. $\rho(H=0)$ plot as observed for newly prepared nc-Ni samples in Ref. 24. These data define a smaller AMR ratio for these old samples than for the new nc-Ni series reported in Ref. 24, the difference being probably due to the sample preparation details and eventual impurity differences.

Further similar experimental work at various temperatures on nc-Ni samples with significantly smaller grain sizes than the current nc-Ni sample might help better understand how the MR($H$) curves and the magnetoresistance anisotropy data are modified by the grain size. Such studies may then provide novel data for analyzing in more detail theoretically the influence of the presence of grain boundaries on the magnetotransport parameters of ferromagnetic metals.

**Acknowledgements**

The Wigner Research Centre for Physics utilizes the research infrastructure of the Hungarian Academy of Sciences (HAS) and is operated by the Eötvös Loránd Research Network (ELKH) Secretariat (Hungary). The authors acknowledge the support of the DFG within the priority program SPP 1666. One of the authors (I.B.) is indebted to the Humboldt Foundation, Germany for a one-month fellowship, to H. Ebert (Ludwig-Maximilians-Universität, München) for the kind hospitality during this research stay as well as to S.T.B. Gönnenwein and R. Gross (Walther-Meissner Institute for Low Temperature Research, Bavarian Academy of Sciences and Humanities, Garching) for generously putting the necessary experimental facility at our disposal in their laboratory for carrying out the resistivity measurements described here. K.P. acknowledges the support by the János Bolyai Research Scholarship of HAS. We are indebted to the Eötvös Loránd University, Budapest, Hungary for providing PhD fellowships to S.Zs. and V.A.I., for the latter via a Stipendium Hungaricum scholarship. The authors also acknowledge L.K. Varga and E. Tóth-Kádár for providing the μc-Ni and nc-Ni samples, respectively.